\documentclass[%
superscriptaddress,
showpacs,
nofootinbib,
amsmath,amssymb,
prc,
twocolumn,
floatfix ]%
{revtex4-1}

\usepackage{color}
\usepackage{CJK}

\usepackage{graphicx}% Include figure files
\usepackage{dcolumn}% Align table columns on decimal point
\usepackage{bm}% bold math
\usepackage{hyperref}% add hypertext capabilities
\allowdisplaybreaks

\begin{document}

\begin{CJK*}{UTF8}{}
\title{Multipole modes of excitation in tetrahedrally deformed neutron-rich Zr isotopes}
\CJKfamily{gbsn}
\author{Jie Zhao (赵杰)}%
%\email{zhaojie@alumni.itp.ac.cn}
\affiliation{Center for Circuits and Systems, Peng Cheng Laboratory, Shenzhen 518055, China}

\date{\today}

\begin{abstract}
The multipole modes of excitation for tetrahedrally deformed neutron-rich 
Zr isotopes are investigated using the quasiparticle finite amplitude method 
based on the covariant density functional theories.
By employing the density-dependent point-coupling covariant density functional theory 
with the parameter set DD-PC1 in the particle-hole channel and a separable pairing interaction of finite range, 
it is observed that a distinct peak emerges at $\omega = 9.0 - 10.0$ MeV 
in the isoscalar quadrupole $K=0$ strength when $\beta_{32}$ distortion is 
considered for $^{110,112}$Zr. 
This peak is absent when the deformation is limited to axially
symmetric octuple or spherical case.
It also does not appears in neighboring axially quadrople or octupole 
deformed nuclei, thus can be viewed as an indicator for the tetrahedral shape.
\end{abstract}

\maketitle

\end{CJK*}

\bigskip

%\section{Introduction~\label{sec:Introduction}}
\emph{Introduction.--}
The occurrence of tetrahedral symmetry is common in nature.
Many quantum objects governed by electromagnetic interactions
such as certain molecules, metal clusters, fullerines, quantum whirlpools, 
and quantum dots can exhibit a tetrahedral shape.
The possibility that atomic nuclei, which are governed by strong interactions 
may have a ground or isometric sate with tetrahedral shape, has also been discussed, 
but still not confirmed experimentally.
A nucleus with a tetrahedral shape is characterized by a pure $\beta_{32}$ deformation 
($\beta_{\lambda\mu}=0$ if $\lambda\ne3$ or $\mu\ne2$).
The corresponding single-particle Hamiltonian has the symmetry group of $T_d^D$,
and the single-particle levels split into multiplets with degeneracies equal to 
the irreducible representations of the $T_d^D$ group.
Due to the fourfold degeneracy in the single-particle levels, 
the predicted shell gaps at specific proton or neutron numbers are comparable 
or even stronger than those at spherical shapes~\cite{Li1994_PRC49-R1250,Dudek2002_PRL88-252502,Dudek2007_IJMPE16-516,
	Dudek2003_APPB34-2491,Heiss1999_PRC60-034303,Arita2014_PRC89-054308}.
Thus, there may be a static tetrahedral shape or strong tetrahedral 
correlations for a nucleus with such proton or neutron numbers.

Many theoretical approaches are used to investigate the possible tetrahedrally 
deformed nuclei,
for example the macroscopic-microscopic model 
\cite{Dudek2002_PRL88-252502,Dudek2007_IJMPE16-516,Dudek2006_PRL97-072501,%
	Schunck2004_PRC69-061305R,Dudek2014_PS89-054007,Jachimowicz2017_PRC95-034329,
	Yang2022_PRC106-054314,Yang2022_PRC105-034348},
the nonrelativistic~\cite{Dudek2007_IJMPE16-516,Schunck2004_PRC69-061305R,Dudek2006_PRL97-072501,%
	Yamagami2001_NPA693-579,Olbratowski2006_IJMPE15-333,Zberecki2009_PRC79-014319,%
	Zberecki2006_PRC74-051302R,Takami1998_PLB431-242,Wang2019_PLB790-498}
and relativistic~\cite{Zhao2017_PRC95-014320,Zhao2024_PRC109-014303,Xu2024_PRC109-014311} 
density functional theories, the algebraic cluster model~\cite{Bijker2014_PRL112-152501}, 
the reflection asymmetric shell model~\cite{Gao2004_CPL21-806,Chen2010_NPA834-378c},
and the $ab$ $initio$ lattice calculation in the framework of nuclear lattice effective field theory~\cite{Epelbaum2014_PRL112-102501}.
The rotational properties of tetrahedral nuclei have also been explored theoretically 
\cite{Gao2004_CPL21-806,Tagami2013_PRC87-054306,Tagami2015_JPG42-015106,
	Chen2010_NPA834-378c,Chen2013_NPR30-278,Bijker2014_PRL112-152501,Tagami2018_PRC98-024304}.
A nucleus with tetrahedral shape is characterized by the occurrence of 
negative-parity bands.
Due to the vanishing quadrupole moments, the rotational bands based on such shapes 
are anticipated to have very weak, or vanishing, 
in-band E2 transitions~\cite{Schunck2004_PRC69-061305R,Bark2010_PRL104-022501,Jentschel2010_PRL104-222502}.
Several experiments have been conducted to identify the tetrahedral shape of 
nuclei based on these criteria.
For instance, the negative-parity bands in $^{160}$Yb and
$^{154,156}$Gd~\cite{Bark2010_PRL104-022501,Jentschel2010_PRL104-222502,Doan2010_PRC82-067306},
$^{230,232}$U~\cite{Ntshangase2010_PRC82-041305R}, 
and $^{156}$Dy~\cite{Hartley2017_PRC95-014321} have been analyzed experimentally, 
but providing evidence against tetrahedral symmetry.

%The negative-parity bands in $^{160}$Yb and $^{154,156}$Gd have been suggested as 
%candidates for rotational bands based on 
%tetrahedral shape~\cite{Dudek2006_PRL97-072501}.
%However, the measured intrinsic quadrupole moments are large, 
%providing strong evidence against tetrahedral %symmetry~\cite{Bark2010_PRL104-022501,Jentschel2010_PRL104-222502,Doan2010_PRC82-067306}.
%By comparing with $^{226}$Ra, the close similarity in the energies 
%and electric dipole moments 
%of the negative-parity bands of $^{230,232}$U implies an 
%octupole vibrational picture rather than 
%the tetrahedral assumption~\cite{Ntshangase2010_PRC82-041305R}.
%An observed long-enough-lived isomer of $^{108}$Zr is expected 
%to have tetrahedral shape~\cite{Sumikama2011_PRL106-202501},
%but the measurement of the corresponding band structure is still missing. 
%The $^{156}$Dy is also predicted to have tetrahedral shape;
%however, the analysis of the $B(E2)/B(E1)$ ratio
%leads to the conclusion that the observed negative-parity bands correspond to 
%octupole excitations rather than the exotic tetrahedral symmetry~\cite{Hartley2017_PRC95-014321}.

Further theoretical investigations suggest that nuclei with tetrahedral symmetry
still can generate sizeable quadrupole transitions due to quadrupole vibrations around 
the equilibrium of pure tetrahedral symmetry~\cite{Dobrowolski2011_IJMPE20-500}. 
Microscopic investigations based on the generator coordinate method even suggest 
that a clear identification of tetrahedral deformations is unlikely due to 
strong mixing with the axial octupole mode~\cite{Zberecki2009_PRC79-014319}.

It is well known that the collective modes exhibited in 
responses to nuclei can provide crucial insights into their shapes.
For instance, the isoscalar giant monopole resonance (ISGMR) splits into two peaks
in prolately deformed nuclei. The lower peak is associated with the mixing 
between the ISGMR and the $K^{\pi} = 0^{+}$ component of the isoscalar giant quadrupole 
resonance (ISGQR), where $K$ denotes the $z$ component of the angular 
momentum~\cite{Garg2018_PPNP101-55,Garg1980_PRL45-1670,Yoshida2010_MPLA25-1783,
	Yoshida2010_PRC82-034324,Kvasil2016_PRC94-064302,Sun2022_PRC105-044312}.
The effect of triaxiality on the multipole excitation of $^{86}$Ge~\cite{Sun2022_PRC106-024334} 
and $^{110}$Ru, $^{190}$Pt~\cite{Washiyama2017_PRC96-041304} has also been studied. 
Recent analysis of the ISGMR in $^{100}$Mo reveals that the appearance of 
a third peak between the two existing peaks can be viewed as an indicator for 
static triaxial deformation~\cite{Shi2023_CPC47-034105,Washiyama2024_PRC109-024317}.

%The Zr isotopes exhibit a rich set of phenomena and open questions, 
%in particular in the areas of shape evolution and shape coexistence~\cite{Chimanski_arXiv2308-13374}.
%The effect of nuclear deformations on the excited-state 
%properties and the multipole strengths of Zr isotopes has already 
%been studied previously within quasiparticle random-phase
%approximation (QRPA) model, but are limited to axial and triaxial quadrupole
%deformations~\cite{Chimanski_arXiv2308-13374,Yoshida2010_PRC82-034324,ElAdri2024_EPJP139-75}.
In Ref.~\cite{Zhao2017_PRC95-014320}, we investigated the potential energy curves of 
even-even neutron-rich Zr isotopes within the multidimensionally constrained relativistic 
Hartree-Bogoliubov (MDC-RHB) model, predicting that the ground state shape 
of $^{110,112}$Zr is tetrahedral.
In this paper, we utilize the RHB plus QRPA model
to explore how tetrahedrally deformed neutron-rich Zr isotopes 
respond to external multipole perturbations.
We explore the finite amplitude 
method (FAM)~\cite{Nakatsukasa2007_PRC76-024318,Avogadro2011_PRC84-014314}
to derive the response function.
Our analysis demonstrates that an additional peak appears at $\omega = 9.0 - 10.0$ MeV 
in the isoscalar quadrupole (ISQ) $K=0$ strength due to $\beta_{32}$ distortion.
This characteristic is unique to the consideration of tetrahedral deformation
and markedly different from neighboring axially quadrople or octupole 
deformed nuclei.
%The theoretical framework and method are introduced in Sec.~\ref{sec:model}.
%The results are described and discussed in Sec~\ref{sec:results}. 
%Sec.~\ref{sec:summary} contains a summary of the principal results. 

%%%%%%%%%%%%%%%%%%%%%%%%%%%%%%%%%%%%%%%%%%%%
%\section{\label{sec:model}Theoretical framework}
%%%%%%%%%%%%%%%%%%%%%%%%%%%%%%%%%%%%%%%%%%%%
\emph{Model.--}
The constrained RHB calculations are preformed with the 
extend MDC-RHB model~\cite{Zhao2024_PRC109-014303}, 
where the RHB equation
\begin{equation}
	\label{eq:rhb}
%	\int d^{3}\bm{r}^{\prime}
	\left(\begin{array}{cc} h_{D}-\lambda & \Delta \\ -\Delta^{*} & -h_{D}+\lambda \end{array}\right)
	\left(\begin{array}{c} U_{k} \\ V_{k} \end{array}\right)
	= E_{k}\left(\begin{array}{c} U_{k} \\ V_{k} \end{array}\right),
\end{equation}
is solved by expanding the Dirac spinors in simplex-$y$ harmonic oscillator (HO) basis.
$E_{k}$ is the quasiparticle energy, 
$\left( U_k, \ V_{k} \right)^\mathrm{T}$ is the quasiparticle wave function,
$\lambda$ is the chemical potential 
and $h_{D}$ is the single-particle Dirac Hamiltonian,
$\Delta$ is the pairing potential.
Our calculations are carried out with the covariant 
density functional DD-PC1~\cite{Niksic2008_PRC77-034302}.
In the pp channel, we use a separable pairing force of finite range \cite{Tian2009_PLB676-44, 
	Tian2009_PRC80-024313, Niksic2010_PRC81-054318}.
%In practical calculations, the ADHO basis is truncated as
%$\left[ n_{z} / Q_{z} + (2n_{\rho}+|m|) / Q_{\rho} \right] \leq N_{f}$
%\cite{Warda2002_PRC66-014310,Lu2014_PRC89-014323},
%for the large component of the Dirac spinor. 
%$N_{f}$ is a certain integer constant and $Q_{z}$ and $Q_{\rho}$ 
%are constants calculated from the oscillator lengths. 
%For the small component, the truncation is made up to $N_g=N_f+1$ major 
%shells in order to avoid spurious states \cite{Gambhir1990_APNY198-132}.
%$N_{f}=20$ is adopted in the present calculations.
Calculations have been performed in a basis with 20 oscillator shells.
The deformation parameter $\beta_{\lambda\mu}$ is obtained from the corresponding
multipole moment using
\begin{equation}
	\beta_{\lambda\mu}^{\tau}=\frac{4\pi}{3N_{\tau}R^{\lambda}}Q_{\lambda\mu}^{\tau},
\end{equation}
where $R=1.2 \times A^{1/3}$ fm
and $N_{\tau}$ is the number of proton, neutron, or nucleons.
%For the details of the RHB model and the ADHO basis, 
%we refer the readers to Ref~\cite{Zhao2017_PRC95-014320}.

The evolution of the nucleonic density subject to a time-dependent 
external perturbation $\hat{F}(t)$ is determined by the 
time-dependent RHB equation.
For a weak harmonic external field, 
the density undergoes small-amplitude oscillations around the equilibrium. 
In the small-amplitude limit, we obtain the FAM equations
\begin{equation}
\label{eq:fam}
\begin{aligned}
	\left( E_{\mu} + E_{\nu} - \omega \right) X_{\mu\nu}(\omega) 
		+ \delta H_{\mu\nu}^{20}(\omega) &= - F_{\mu\nu}^{20}, \\
		\left( E_{\mu} + E_{\nu} + \omega \right) Y_{\mu\nu}(\omega) 
	+ \delta H_{\mu\nu}^{02}(\omega) &= - F_{\mu\nu}^{02},
\end{aligned}
\end{equation}
where $X$ and $Y$ are FAM amplitudes at a given frequency $\omega$. 
$\delta H^{20}$ ($\delta H^{02}$) and $F^{20}$ ($F^{02}$) are two-quasiparticle matrix elements of an 
induced Hamiltonian and an external field respectively. 
%The FAM equations Eq.~(\ref{eq:fam}) can be solved iteratively.
We solve the FAM equations iteratively in simplex-$y$ HO basis 
with 20 oscillator shells, same as the case of solving the static RHB equation (\ref{eq:rhb}).
The imaginary part of the frequency $\omega$ has been introduced as 
$\omega \rightarrow \omega + i\gamma$ with $\gamma = 1.0$ MeV. 
The spacing in discretized $\omega$ is taken to be $0.5$ MeV to compute strength functions 
${dB(f,\omega)}/{d\omega}$.
%\begin{equation}
%	\frac{dB(f,\omega)}{d\omega} = -\frac{1}{\pi} {\rm{Im}} S(f,\omega)
%				= -\frac{1}{\pi} {\rm{Im}} {\rm{Tr}} \left[ f^{\dagger} 
%				  \delta \rho(\omega) \right],
%\end{equation}
%where $\delta \rho (\omega)$ denotes the induced density matrix 
%and $f_{kl}$ are the matrix elements of the operator $F(\omega)$ in the 
%single-particle basis.
The transition density for each particular frequency $\omega$ is defined as 
\begin{equation}
	\delta \rho_{\rm{tr}} (\bm{r}) = -\frac{1}{\pi} {\rm{Im}} \delta \rho(\bm{r}).
\end{equation}
The isoscalar and isovector multipole operators are defined as 
\begin{equation}
	f^{IS}_{JK} = \sum_{i=1}^{A} f_{JK}^{(+)} (\bm{r}_{i}), \quad
	f^{IV}_{JK} = \sum_{\tau} 
		\sum_{i=1}^{N_{\tau}} (-\tau) f_{JK}^{(+)} (\bm{r}_{i}),
\end{equation}
where we define $f_{JK}^{(+)} = \left[ f_{JK} + (-1)^{K} f_{J-K} \right] / \sqrt{2+2\delta_{K0}}$
with $f_{JK} (\bm{r}) = r^{J} Y_{JK} (\theta,\varphi)$ for $J \neq 0$ and $f_{00}(\bm{r}) = r^{2}$.
$\tau = +1$ ($-1$) refers to neutron (proton).
For the isovector dipole excitation ($D_{k} = r Y_{1k}$, $k=0, \pm 1$), 
we adopt the definition~\cite{Bjelcic2020_CPC253-107184}
\begin{equation}
D_{k} = e \frac{NZ}{A} \left[ \frac{1}{Z} \sum_{i=1}^{Z} D_{k}(\bm{r}_{i}) 
							- \frac{1}{N} \sum_{i=1}^{N} D_{k}(\bm{r}_{i}) \right].
\end{equation}
The translational spurious modes are removed following the prescription 
proposed in Ref.~\cite{Nakatsukasa2007_PRC76-024318}.
For the details of the formulas of the relativistic QFAM, 
we refer the readers to Ref.~\cite{Bjelcic2020_CPC253-107184}.
The discrete QRPA transition strength $|\langle i | \hat{F} | 0 \rangle |^{2}$
and the eigenfrequencies $\Omega_{i}$
can be extracted from the QFAM calculation via the contour integration in the 
complex plane~\cite{Hinohara2013_PRC87-064309}.
%where
%\begin{equation}
%|\langle i | \hat{F} | 0 \rangle |^{2} = \frac{1}{2\pi i} 
%	\oint_{C_i} S(f,\omega) d\omega
%\label{eq:tran}
%\end{equation}
%and 
%\begin{equation}
%\begin{aligned}
%\Omega_{i}^{2} =& \frac{1}{|\langle i | \hat{F} | 0 \rangle |^{2}} 
%		\sum_{\mu<\nu}  \\
%	 & \left\{ 
%		\left| \frac{1}{2\pi i} \oint_{C_i} \left[ (E_{\mu} + E_{\nu})
%			    X_{\mu\nu}(\omega) 
%			  + \delta H_{\mu\nu}^{20} (\omega) \right]
%			  d\omega \right|^{2} \right. \\
%	 &- \left. \left| \frac{1}{2\pi i} \oint_{C_i} \left[ (E_{\mu} + E_{\nu})
%	    Y_{\mu\nu}(\omega) 
%	    + \delta H_{\mu\nu}^{02} (\omega) \right] 
%	    d\omega \right|^{2}
%			  \right\}.
%\end{aligned}
%\label{eq:omeg}
%\end{equation}
The closed contour $C_{i}$ in the complex energy plane encloses the $i$-th
positive pole $\Omega_{i}$.
In principle, we can select any closed simple loop $C_{i}$,
here, we choose a circle
$\omega (\varphi) = \omega_{0} + \omega_{R} e^{i\varphi}$,
and the center $\omega_{0}$ and the radius $\omega_{R}$ is estimated from the response function profile in order that the contour encircles the desired pole $\Omega_{i}$.
We set $\omega_{R} = 0.05$ MeV and 80 quadrature points are used to evaluate
the integral numerically.

\emph{Results and discussions.--}
In Ref.~\cite{Zhao2017_PRC95-014320}, we presented the potential energy curve of $^{110}$Zr.
Imposing axial and reflection symmetry predicts
the ground state shape to be oblate with $\beta_{20} \approx -0.2$. 
Additionally, two additional minima are observed at $\beta_{20} \approx 0$ and $0.5$.
However, the spherical minimum is unstable with respect to octupole deformations.
The inclusion of both $\beta_{30}$ and $\beta_{32}$ can lowering its energy,
with the effect being much stronger for $\beta_{32}$ than for $\beta_{30}$.
Consequently, the ground state shape of $^{110}$Zr is predicted 
to be tetrahedral with $\beta_{32} = 0.18$, 
and a pear-shaped minimum ($\beta_{30}=0.20$) with energy approximately $1$ MeV above
the tetrahedrally deformed ground state minimum can also be observed.

\begin{figure}[htb]
	\begin{center}
		\includegraphics[width=0.4\textwidth]{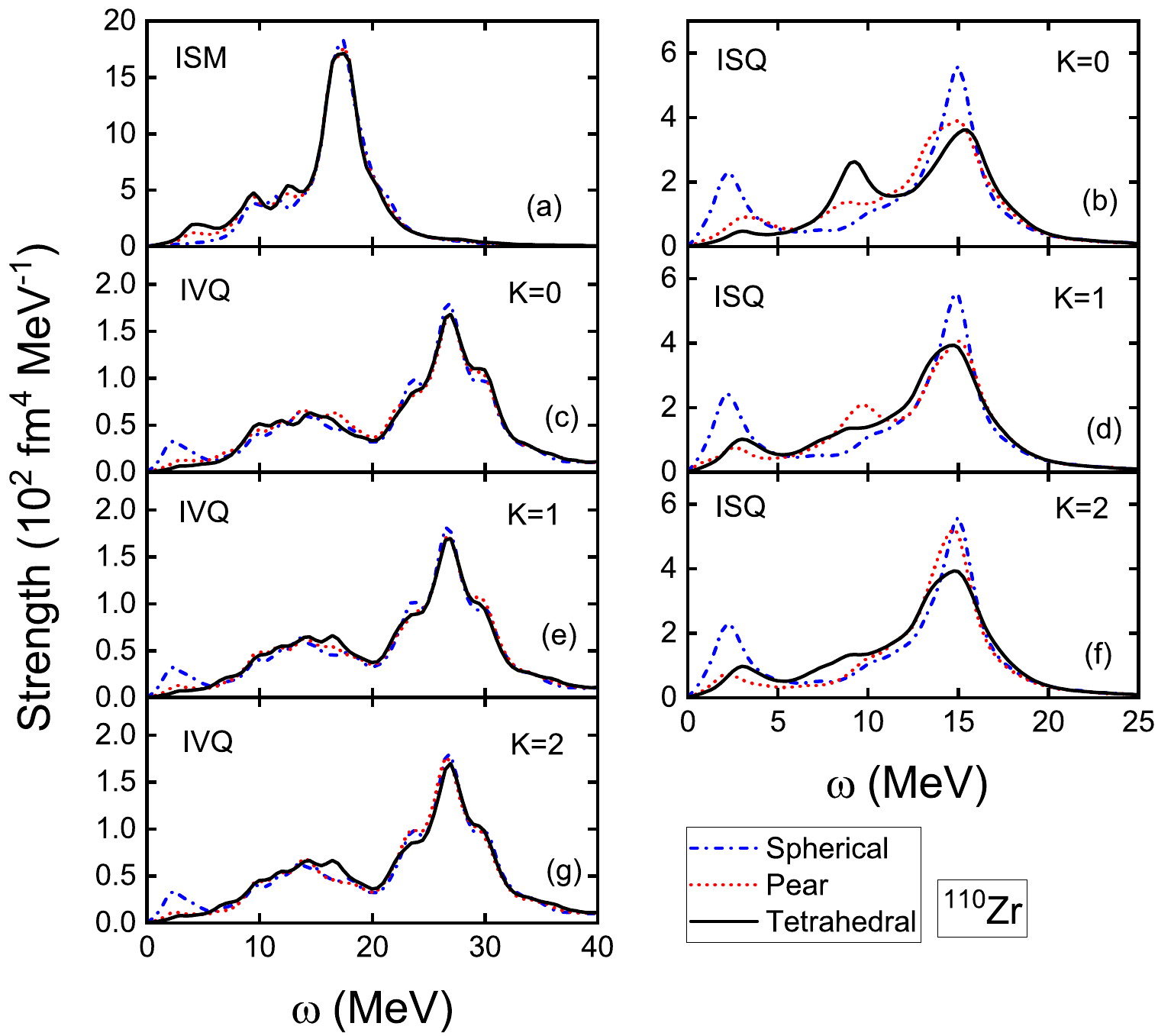}
	\end{center}
	\caption{(Color online) 
		Isoscalar monopole (a), isoscalar (b,d,f) and isovector (c,e,g) quadrupole strength 
		as a function of $\omega$ for $^{110}$Zr.
		The dash-dotted (dotted) line denotes the results obtained 
		at the point with spherical (pear) shape.
		The results obtained at the ground state 
		minimum with tetrahedral shape are represented by solid line.
	}
	\label{fig:Multi}
\end{figure}

To investigate the response of the tetrahedrally deformed nucleus 
$^{110}$Zr to isoscalar and isovector multipole perturbations, 
we computed the monopole and quadrupole transition strengths using FAM. 
The results are depicted in Fig.~\ref{fig:Multi} (solid line). 
For comparison, results calculated at the spherical shape 
and at the pear-shaped minimum are also presented 
with dash-dotted and dotted lines, respectively.
In Fig.~\ref{fig:Multi} (a), the isoscalar monopole (ISM) strength is displayed. 
Generally, the ISM strength calculated at these three deformation points is similar.
A giant resonance peak appears at $\omega \approx 17.5$ MeV in all cases, 
with only slight differences around $\omega \approx 4.5$, $9.5$, and $12.5$ MeV.
The obtained isovector quadrupole (IVQ) $K = 0, 1, 2$ strengths are shown in 
Fig.~\ref{fig:Multi} (c), (e), and (g) respectively.
The IVQ strength for different $K$ calculated at the spherical shape agrees with each other, 
with the dominant peak located at $\omega \approx 27$ MeV. 
A small peak at $\omega \approx 2.5$ MeV is also observed in this case.
When calculating the IVQ strength at the pear-shaped minimum, 
this lowest peak disappears,
while the other parts remain similar to those calculated at the spherical shape.
The main structure of the IVQ strength obtained at the tetrahedrally deformed 
ground state minimum resembles that calculated at the pear-shaped minimum.

In Fig.~\ref{fig:Multi} (b), (d), and (f), we present the ISQ
$K = 0, 1, 2$ strengths calculated at the spherical shape, and at the pear and 
tetrahedral shaped minima.
Due to spherical symmetry, the ISQ strengths calculated at the spherical deformation point 
are identical for different $K$ and exhibit a two-peak structure. 
The smaller peak is located at $\omega \approx 2.0$ MeV, while the dominant 
peak is located at $\omega \approx 15.0$ MeV.
When calculating the ISQ strengths at the pear-shaped minimum, 
the position of the smaller peak shifts to $\omega \approx 3.0$ MeV, 
and its magnitude decreases significantly. 
For $K=0$, a shoulder appears around $\omega \approx 9.0$ MeV, 
and a very small peak is observed at this position for $K=1$.
Similarly, when calculating the ISQ strengths at the ground state minimum 
with tetrahedral shape, for $K=0$, in addition to the giant resonance peak 
at $\omega \approx 15.0$ MeV and the small peak at $\omega \approx 3.0$ MeV,
another significant peak emerges at $\omega \approx 9.0$ MeV. 
The structure of the ISQ strengths for $K=1$ and $2$ remains similar, 
with only the dominant peak at $\omega \approx 15.0$ MeV and the small peak 
at $\omega \approx 3.0$ MeV observed.
Thus, the appearance of the peak at $\omega \approx 9.0$ MeV is unique to the ISQ $K=0$ strength.

\begin{figure}[htb]
	\begin{center}
		\includegraphics[width=0.4\textwidth]{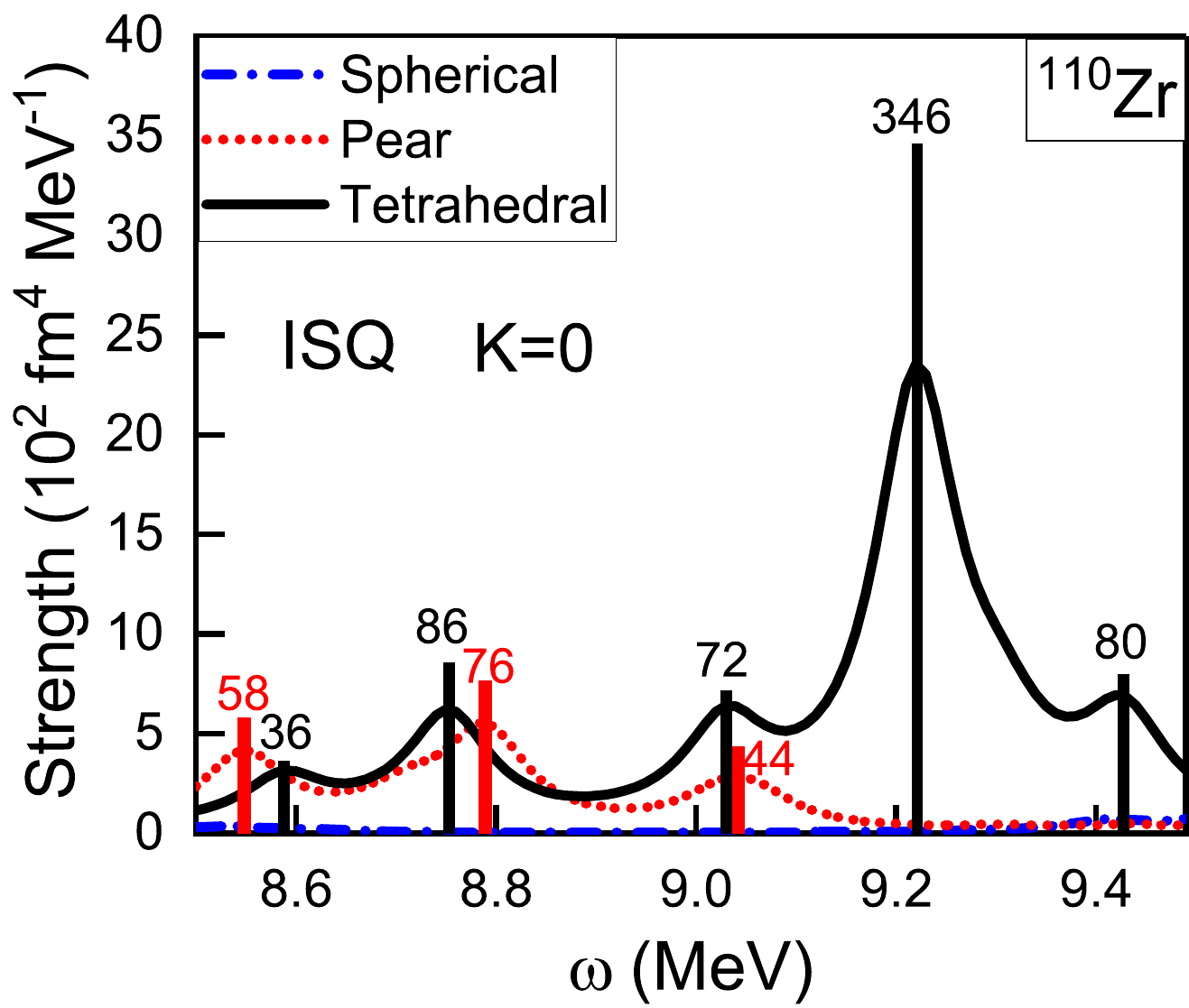}
	\end{center}
	\caption{(Color online) 
		Isoscalar quadrupole $K=0$ strength at $\omega = 8.5-9.5$ MeV for $^{110}$Zr
		calculated with $\gamma=0.05$ MeV.
		The dash-dotted (dotted) line denotes the results obtained at the point with 
		spherical (pear) shape.
		The results obtained at the ground state minimum with tetrahedral shape
		are represented by solid line.
		The bars denote the QRPA eigenfrequencies obtained via the 
		method based on the contour integration in the complex plane, 
		and the corresponding numbers denote the discrete QRPA transition 
		strength (in fm$^{4}$). The red (black) ones denote the results calculated
		at the minimum with pear shape (tetrahedral shape).
	}
	\label{fig:Level}
\end{figure}

Let's focus on the ISQ $K=0$ strength.
As depicted in Fig.~\ref{fig:Multi} (b),
the peak at $\omega \approx 9.0$ MeV only emerges when including $\beta_{32}$ deformation. 
To gain further insight into the origin of this peak, 
we utilize the method based on contour integration in the complex 
plane~\cite{Hinohara2013_PRC87-064309} to extract the QRPA 
transition matrix elements and eigenfrequencies from the QFAM calculation.
Initially, we run the QFAM calculations with $\gamma = 0.05$ MeV,
obtaining a response profile $dB(f,\omega)/d\omega$ that provides 
a reasonable estimate for the location of the QRPA poles $\Omega_{i}$ 
with significant transition probabilities $|\langle i | \hat{F} | 0 \rangle |^{2}$, 
as illustrated in Fig.~\ref{fig:Level}. 
For the spherical case, no significant peaks appear in the range $\omega = 8.5-9.5$ MeV.
When calculated at the pear-shaped minimum, 
three small peaks are observed around $\omega \approx 8.55, 8.79$, and $9.04$ MeV.
Similarly, at the tetrahedrally deformed ground state minimum, 
in addition to the three peaks around $\omega \approx 8.59, 8.75$, and $9.03$ MeV 
similar to those at the pear-shaped minimum, 
two additional peaks are observed around $\omega \approx 9.22$ and $9.42$ MeV.
Next, a contour integration is performed to calculate the discrete QRPA 
transition strength and the eigenfrequencies.
% [Eqs.~(\ref{eq:tran}) and (\ref{eq:omeg})]. 
At the pear-shaped minimum, 
the QRPA transition strengths for $\Omega_{i} = 8.55, 8.79$, and $9.04$ MeV 
are 58, 76, and 44 fm$^{4}$, respectively. 
For the tetrahedrally deformed ground state minimum, 
the QRPA transition strengths for $\Omega_{i} = 8.59, 8.75, 9.03, 9.22$, 
and $9.43$ MeV are 36, 86, 72, 346, and 80 fm$^{4}$, respectively.
Clearly, the state with $\Omega_{i} = 9.22$ MeV,
which only appears at the tetrahedrally deformed minimum, 
has the largest transition strength, 
contributing the most to the strong peak observed in the ISQ $K=0$ 
strength at $\omega \approx 9.0$ MeV.

\begin{figure}[h]
	\begin{center}
		\includegraphics[width=0.4\textwidth]{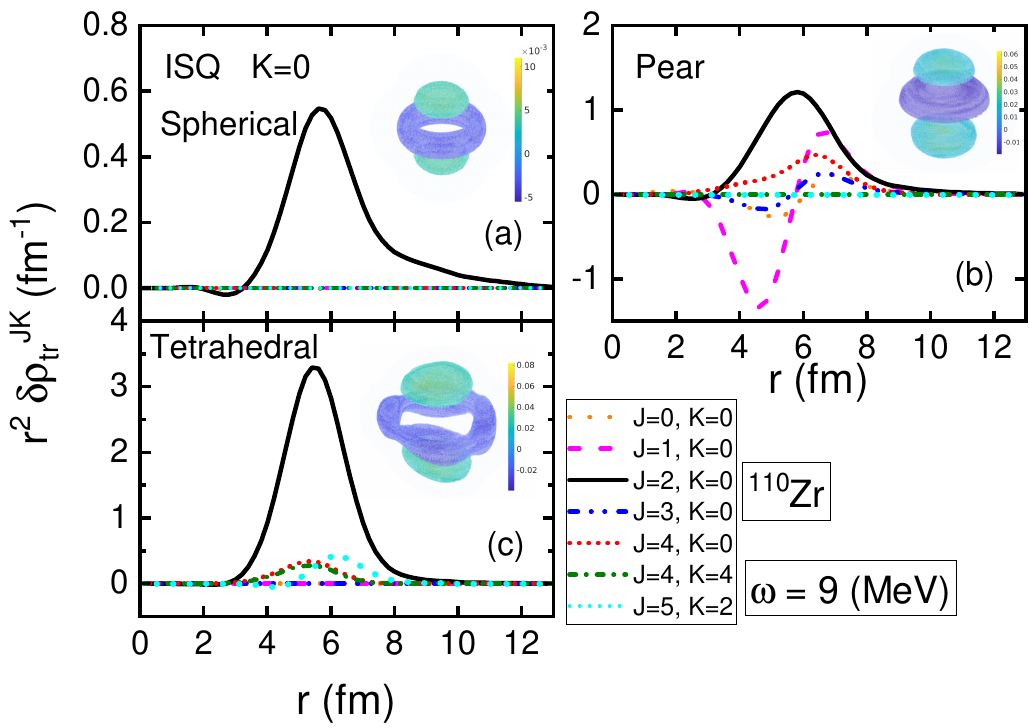}
	\end{center}
	\caption{(Color online) 
		Radial parts of the angular-momentum-projected transition densities
		correspond to the ISQ $K=0$ strength at $\omega=9$ MeV for $^{110}$Zr
		at the point with spherical shape (a), at the pear-shaped minimum (b)
		and at the tetrahedrally deformed ground state minimum (c).
		The corresponding three dimensional transition densities with
		$|\delta \rho_{\rm{tr}}| \geq 3\times10^{-3}$ fm$^{-3}$ (a),
		$|\delta \rho_{\rm{tr}}| \geq 1.4\times10^{-2}$ fm$^{-3}$ (b)
		and $|\delta \rho_{\rm{tr}}| \geq 2\times10^{-2}$ fm$^{-3}$ (c)
		are shown in the inset.
	}
	\label{fig:trDen}
\end{figure}
To gain a deeper understanding of the spatial structure of the induced transition density,
we plotted the radial part of the projected transition density~\cite{Niksic2013_PRC88-044327} 
at $\omega \approx 9.0$ MeV:
\begin{equation}
\delta \rho_{\rm{tr}}^{JK} (r) = \int d\Omega \delta \rho_{\rm{tr}} (r, \Omega) Y_{JK} (\Omega),
\end{equation}
as shown in Fig.~\ref{fig:trDen}. 
Figures \ref{fig:trDen} (a), (b), and (c) display the transition density obtained
at the spherical deformation point, the pear-shaped minimum, and the ground state 
with tetrahedral shape, respectively. 
The corresponding three-dimensional (3D) plots of the induced transition densities 
are also shown in the inset.
Given that the predicted radius of $^{110}$Zr is about 4.7 fm,
the radial densities are concentrated on the nucleus's surface, 
with minimal compensating density change at smaller radii.
At the spherical deformation point [see inset of Fig.~\ref{fig:trDen} (a)], 
the transition density is axial and reflection symmetric, 
with only the $\delta \rho_{\text{tr}}^{20} (r)$ component visible.
When including axial octupole deformation $\beta_{30}$ [Fig.~\ref{fig:trDen} (b)], 
the transition density remains axially symmetric, but reflection symmetry is broken. 
In addition to the dominant $\delta \rho_{\text{tr}}^{20} (r)$ component,
many other components including $\delta \rho_{\text{tr}}^{00} (r)$, 
$\delta \rho_{\text{tr}}^{10} (r)$, $\delta \rho_{\text{tr}}^{30} (r)$, 
and $\delta \rho_{\text{tr}}^{40} (r)$ become visible.
At the ground state minimum with tetrahedral shape [inset of Fig.~\ref{fig:trDen} (c)],
the transition densities break both axial and reflection symmetry. 
While the dominant component remains $\delta \rho_{\text{tr}}^{20} (r)$, 
the $\delta \rho_{\text{tr}}^{10} (r)$ and $\delta \rho_{\text{tr}}^{30} (r)$ 
components are exactly zero. 
Additionally, the $\delta \rho_{\text{tr}}^{00} (r)$ and $\delta \rho_{\text{tr}}^{32} (r)$ 
components are negligible. 
Instead, the $\delta \rho_{\text{tr}}^{40} (r)$, $\delta \rho_{\text{tr}}^{44} (r)$ 
and $\delta \rho_{\text{tr}}^{52} (r)$
components are significant, with the $\delta \rho_{\text{tr}}^{44} (r)$ 
and $\delta \rho_{\text{tr}}^{52} (r)$ components appearing 
only when considering $\beta_{32}$ deformation.

\begin{figure}[h]
	\begin{center}
		\includegraphics[width=0.4\textwidth]{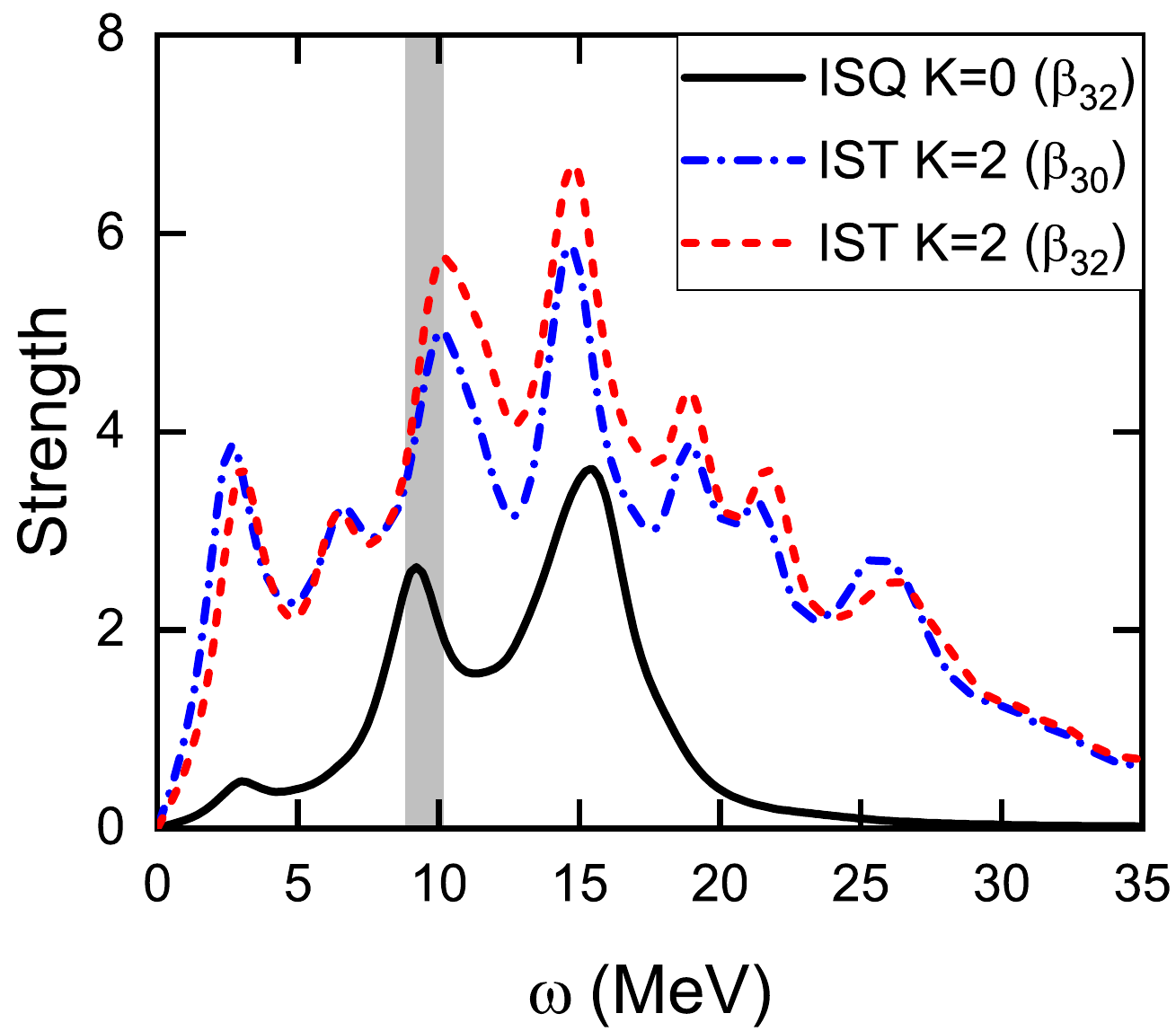}
	\end{center}
	\caption{(Color online)
		Isoscalar triacontadipole strength (in $5\times10^{6}$ fm$^{10}$ MeV$^{-1}$) 
		as a function of $\omega$ for $^{110}$Zr. 
		The dash-dotted (dashed) line denotes the results obtained at the point 
		with pear (tetrahedral) shape. 
		The isoscalar quadrupole $K=0$ strength (in $10^{2}$ fm$^{4}$ MeV$^{-1}$) 
		obtained at the ground state minimum 
		with tetrahedral shape is also shown for comparison (solid line).
	}
	\label{fig:zrISH}
\end{figure}

Accordingly, we calculated the isoscalar triacontadipole (IST) $K=2$ strength.
The obtained results are shown in Fig.~\ref{fig:zrISH}.
In general, the structure of the IST $K=2$ strengths calculated at the pear-shaped
and tetrahedral-shaped minima are similar. 
Several peaks can be observed, located 
at $\omega \approx$ 3.0, 6.5, 10.0, 15.0, 19.0, 21.5, and 26.0 MeV.
The dominant peak is located at $\omega \approx 15.0$ MeV,
and the second dominant one is located at $\omega \approx 10.0$ MeV.
Note that the position of this peak is very close to the one observed 
at $\omega \approx 9.0$ MeV 
for the ISQ $K=0$ strength obtained at the tetrahedral minimum.
This indicates that when $\beta_{32}$ distortion is included, 
the emergence of the peak at $\omega \approx 9.0$ MeV for the ISQ $K=0$ strength  
may be due to the coupling with the $K=2$ component of the IST mode.

\begin{figure}[h]
	\begin{center}
		\includegraphics[width=0.4\textwidth]{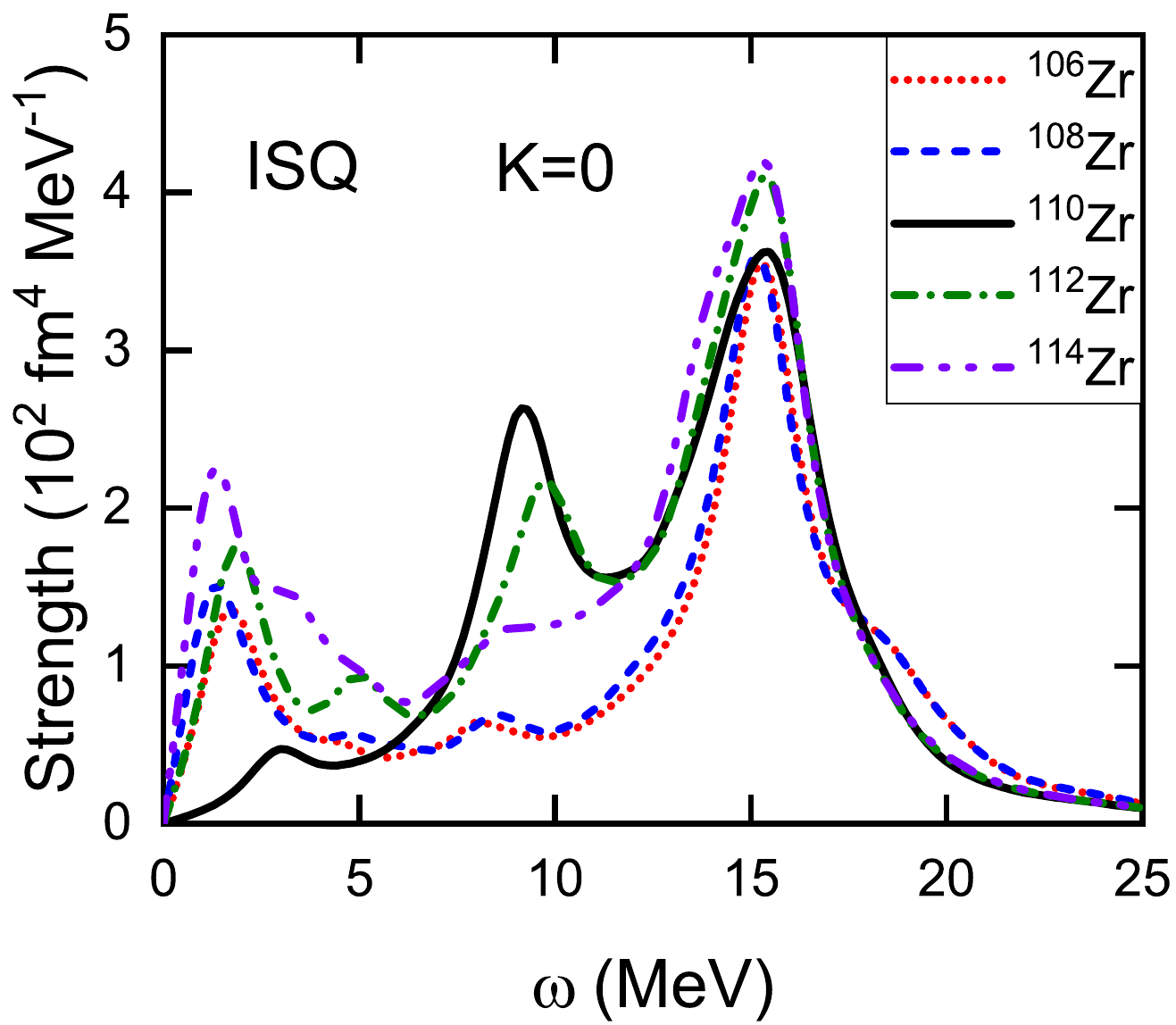}
	\end{center}
	\caption{(Color online)
		Isoscalar quadrupole $K=0$ strength as a function of $\omega$ for $^{106-114}$Zr calculated
		at the corresponding ground state minimum. 
	}
	\label{fig:zrIso}
\end{figure}

In Fig.~\ref{fig:zrIso}, we illustrate the ISQ $K=0$ strengths for even-even neutron-rich isotopes $^{106-114}$Zr, calculated at their respective ground state minima.
For $^{106}$Zr, the ground state shape is predicted to be oblate with $\beta_{20} = -0.23$, 
depicted by the red dotted line. The ISQ $K=0$ strength exhibits a giant resonance peak around $15$ MeV, alongside a lower peak at approximately $\omega \approx 1.5$ MeV.
Similarly, the ground state shape of $^{108}$Zr also features an oblate configuration 
with $\beta_{20} = -0.23$, resulting in an ISQ $K=0$ strength (dashed line) similar to 
that of $^{106}$Zr, as shown in Fig.~\ref{fig:zrIso}.
In contrast, $^{112}$Zr is predicted to possess a tetrahedral shape with $\beta_{32} = 0.15$. 
The corresponding ISQ $K=0$ strength (dash-dotted line) in Fig.~\ref{fig:zrIso} 
exhibits a similar structure to that of $^{110}$Zr obtained at the ground state minimum 
with a tetrahedral shape (solid line), featuring a significant peak at $\omega \approx 10.0$ MeV, 
alongside the lowest peak at $\omega \approx 2.0$ MeV 
and the giant resonance peak at $\omega \approx 15.0$ MeV.
Lastly, for $^{114}$Zr, the ground state is predicted to have a pear shape with $\beta_{30} = 0.15$, resulting in an ISQ $K=0$ strength (dash-dot-dotted line) with a two-peak structure, 
similar to the cases of $^{106,108}$Zr, 
with peaks located at $\omega \approx 1.5$ and $15$ MeV, respectively.
These results suggest that the intermediate peak observed at $\omega \approx 9.0-10.0$ MeV 
in the ISQ $K=0$ strength is specific to $^{110,112}$Zr, 
serving as an indicator for tetrahedral shape.

%
%\section{\label{sec:summary}Summary}
\emph{Summary.--}
In summary, we investigated the multipole modes of excitation for tetrahedrally deformed neutron-rich Zr isotopes using the quasiparticle finite amplitude method based on the covariant density functional DD-PC1 and a separable pairing force. The inclusion of $\beta_{32}$ distortion led to the emergence of a distinctive peak at $\omega \approx 9.0$ MeV in the ISQ $K=0$ strength for $^{110}$Zr. Notably, this peak was absent in cases where the system was restricted to axially symmetric or spherical configurations.
A detailed analysis of the discrete QRPA modes within the range $\omega = 8.5 - 9.5$ MeV revealed significant insights. Specifically, when compared with the case of the pear-shaped minimum, two additional discrete eigenfrequencies, $\Omega_{i} = 9.22$ and 9.43 MeV, appeared at the tetrahedrally deformed ground state minimum. Remarkably, the state with $\Omega_{i} = 9.22$ MeV exhibited the largest transition strength, thereby contributing most to the pronounced peak observed at $\omega \approx 9.0$ MeV.
This unique peak serves as a distinctive feature of the ground states of $^{110,112}$Zr with tetrahedral shapes and is not observed in neighboring axially quadrupole or octupole deformed nuclei. 

%%%%%%%%%%%%%%%%%%%%%%%%%%%%%%%%%%%%%%%%%%
%\bigskip
%---------------------------------------------------------
\begin{acknowledgments}
%\Acknowledgements
%\emph{Acknowledgements.--}
This work has been supported by the National Natural Science 
Foundation of China under Grant No. 12005107.
\end{acknowledgments}

%\bibliographystyle{apsrev4-1}
%\bibliography{nuclear,Nuclear-Fission,MyOwn}
%\bibliography{/home/zhaoj/MyData/MyWork/Paper/Nuclear-Fission/Nuclear-Fission,/home/zhaoj/MyData/MyWork/Paper/Nuclear-Phys/nuclear,/home/zhaoj/MyData/MyWork/Paper/MyOwn/MyOwn}

\begin{thebibliography}{54}%
	\makeatletter
	\providecommand \@ifxundefined [1]{%
		\@ifx{#1\undefined}
	}%
	\providecommand \@ifnum [1]{%
		\ifnum #1\expandafter \@firstoftwo
		\else \expandafter \@secondoftwo
		\fi
	}%
	\providecommand \@ifx [1]{%
		\ifx #1\expandafter \@firstoftwo
		\else \expandafter \@secondoftwo
		\fi
	}%
	\providecommand \natexlab [1]{#1}%
	\providecommand \enquote  [1]{``#1''}%
	\providecommand \bibnamefont  [1]{#1}%
	\providecommand \bibfnamefont [1]{#1}%
	\providecommand \citenamefont [1]{#1}%
	\providecommand \href@noop [0]{\@secondoftwo}%
	\providecommand \href [0]{\begingroup \@sanitize@url \@href}%
	\providecommand \@href[1]{\@@startlink{#1}\@@href}%
	\providecommand \@@href[1]{\endgroup#1\@@endlink}%
	\providecommand \@sanitize@url [0]{\catcode `\\12\catcode `\$12\catcode
		`\&12\catcode `\#12\catcode `\^12\catcode `\_12\catcode `\%12\relax}%
	\providecommand \@@startlink[1]{}%
	\providecommand \@@endlink[0]{}%
	\providecommand \url  [0]{\begingroup\@sanitize@url \@url }%
	\providecommand \@url [1]{\endgroup\@href {#1}{\urlprefix }}%
	\providecommand \urlprefix  [0]{URL }%
	\providecommand \Eprint [0]{\href }%
	\providecommand \doibase [0]{http://dx.doi.org/}%
	\providecommand \selectlanguage [0]{\@gobble}%
	\providecommand \bibinfo  [0]{\@secondoftwo}%
	\providecommand \bibfield  [0]{\@secondoftwo}%
	\providecommand \translation [1]{[#1]}%
	\providecommand \BibitemOpen [0]{}%
	\providecommand \bibitemStop [0]{}%
	\providecommand \bibitemNoStop [0]{.\EOS\space}%
	\providecommand \EOS [0]{\spacefactor3000\relax}%
	\providecommand \BibitemShut  [1]{\csname bibitem#1\endcsname}%
	\let\auto@bib@innerbib\@empty
	%</preamble>
	\bibitem [{\citenamefont {Li}\ and\ \citenamefont
		{Dudek}(1994)}]{Li1994_PRC49-R1250}%
	\BibitemOpen
	\bibfield  {author} {\bibinfo {author} {\bibfnamefont {X.}~\bibnamefont
			{Li}}\ and\ \bibinfo {author} {\bibfnamefont {J.}~\bibnamefont {Dudek}},\
	}\href {http://link.aps.org/doi/10.1103/PhysRevC.49.R1250} {\bibfield
		{journal} {\bibinfo  {journal} {Phys. Rev. C}\ }\textbf {\bibinfo {volume}
			{49}},\ \bibinfo {pages} {R1250} (\bibinfo {year} {1994})}\BibitemShut
	{NoStop}%
	\bibitem [{\citenamefont {Dudek}\ \emph {et~al.}(2002)\citenamefont {Dudek},
		\citenamefont {G\'o\'zd\'z}, \citenamefont {Schunck},\ and\ \citenamefont
		{Mi\'skiewicz}}]{Dudek2002_PRL88-252502}%
	\BibitemOpen
	\bibfield  {author} {\bibinfo {author} {\bibfnamefont {J.}~\bibnamefont
			{Dudek}}, \bibinfo {author} {\bibfnamefont {A.}~\bibnamefont {G\'o\'zd\'z}},
		\bibinfo {author} {\bibfnamefont {N.}~\bibnamefont {Schunck}}, \ and\
		\bibinfo {author} {\bibfnamefont {M.}~\bibnamefont {Mi\'skiewicz}},\ }\href
	{http://link.aps.org/doi/10.1103/PhysRevLett.88.252502} {\bibfield  {journal}
		{\bibinfo  {journal} {Phys. Rev. Lett.}\ }\textbf {\bibinfo {volume} {88}},\
		\bibinfo {pages} {252502} (\bibinfo {year} {2002})}\BibitemShut {NoStop}%
	\bibitem [{\citenamefont {Dudek}\ \emph {et~al.}(2007)\citenamefont {Dudek},
		\citenamefont {Dobaczewski}, \citenamefont {Dubray}, \citenamefont {Pangon},\
		and\ \citenamefont {Schunck}}]{Dudek2007_IJMPE16-516}%
	\BibitemOpen
	\bibfield  {author} {\bibinfo {author} {\bibfnamefont {J.}~\bibnamefont
			{Dudek}}, \bibinfo {author} {\bibfnamefont {J.}~\bibnamefont {Dobaczewski}},
		\bibinfo {author} {\bibfnamefont {A.}~\bibnamefont {Dubray}, \bibfnamefont
			{N.and~G\'o\'zd\'z}}, \bibinfo {author} {\bibfnamefont {V.}~\bibnamefont
			{Pangon}}, \ and\ \bibinfo {author} {\bibfnamefont {N.}~\bibnamefont
			{Schunck}},\ }\href {\doibase 10.1142/S0218301307005958} {\bibfield
		{journal} {\bibinfo  {journal} {Int. J. Mod. Phys. E}\ }\textbf {\bibinfo
			{volume} {16}},\ \bibinfo {pages} {516} (\bibinfo {year} {2007})}\BibitemShut
	{NoStop}%
	\bibitem [{\citenamefont {Dudek}\ \emph {et~al.}(2003)\citenamefont {Dudek},
		\citenamefont {G\'o\'zd\'z},\ and\ \citenamefont
		{Schunck}}]{Dudek2003_APPB34-2491}%
	\BibitemOpen
	\bibfield  {author} {\bibinfo {author} {\bibfnamefont {J.}~\bibnamefont
			{Dudek}}, \bibinfo {author} {\bibfnamefont {A.}~\bibnamefont {G\'o\'zd\'z}},
		\ and\ \bibinfo {author} {\bibfnamefont {N.}~\bibnamefont {Schunck}},\ }\href
	{https://www.actaphys.uj.edu.pl/R/34/4/2491} {\bibfield  {journal} {\bibinfo
			{journal} {Acta Phys. Pol. B}\ }\textbf {\bibinfo {volume} {34}},\ \bibinfo
		{pages} {2491} (\bibinfo {year} {2003})}\BibitemShut {NoStop}%
	\bibitem [{\citenamefont {Heiss}\ \emph {et~al.}(1999)\citenamefont {Heiss},
		\citenamefont {Lynch},\ and\ \citenamefont
		{Nazmitdinov}}]{Heiss1999_PRC60-034303}%
	\BibitemOpen
	\bibfield  {author} {\bibinfo {author} {\bibfnamefont {W.~D.}\ \bibnamefont
			{Heiss}}, \bibinfo {author} {\bibfnamefont {R.~A.}\ \bibnamefont {Lynch}}, \
		and\ \bibinfo {author} {\bibfnamefont {R.~G.}\ \bibnamefont {Nazmitdinov}},\
	}\href {http://link.aps.org/doi/10.1103/PhysRevC.60.034303} {\bibfield
		{journal} {\bibinfo  {journal} {Phys. Rev. C}\ }\textbf {\bibinfo {volume}
			{60}},\ \bibinfo {pages} {034303} (\bibinfo {year} {1999})}\BibitemShut
	{NoStop}%
	\bibitem [{\citenamefont {Arita}\ and\ \citenamefont
		{Mukumoto}(2014)}]{Arita2014_PRC89-054308}%
	\BibitemOpen
	\bibfield  {author} {\bibinfo {author} {\bibfnamefont {K.-i.}\ \bibnamefont
			{Arita}}\ and\ \bibinfo {author} {\bibfnamefont {Y.}~\bibnamefont
			{Mukumoto}},\ }\href {http://link.aps.org/doi/10.1103/PhysRevC.89.054308}
	{\bibfield  {journal} {\bibinfo  {journal} {Phys. Rev. C}\ }\textbf {\bibinfo
			{volume} {89}},\ \bibinfo {pages} {054308} (\bibinfo {year}
		{2014})}\BibitemShut {NoStop}%
	\bibitem [{\citenamefont {Dudek}\ \emph {et~al.}(2006)\citenamefont {Dudek},
		\citenamefont {Curien}, \citenamefont {Dubray}, \citenamefont {Dobaczewski},
		\citenamefont {Pangon}, \citenamefont {Olbratowski},\ and\ \citenamefont
		{Schunck}}]{Dudek2006_PRL97-072501}%
	\BibitemOpen
	\bibfield  {author} {\bibinfo {author} {\bibfnamefont {J.}~\bibnamefont
			{Dudek}}, \bibinfo {author} {\bibfnamefont {D.}~\bibnamefont {Curien}},
		\bibinfo {author} {\bibfnamefont {N.}~\bibnamefont {Dubray}}, \bibinfo
		{author} {\bibfnamefont {J.}~\bibnamefont {Dobaczewski}}, \bibinfo {author}
		{\bibfnamefont {V.}~\bibnamefont {Pangon}}, \bibinfo {author} {\bibfnamefont
			{P.}~\bibnamefont {Olbratowski}}, \ and\ \bibinfo {author} {\bibfnamefont
			{N.}~\bibnamefont {Schunck}},\ }\href
	{http://link.aps.org/doi/10.1103/PhysRevLett.97.072501} {\bibfield  {journal}
		{\bibinfo  {journal} {Phys. Rev. Lett.}\ }\textbf {\bibinfo {volume} {97}},\
		\bibinfo {pages} {072501} (\bibinfo {year} {2006})}\BibitemShut {NoStop}%
	\bibitem [{\citenamefont {Schunck}\ \emph {et~al.}(2004)\citenamefont
		{Schunck}, \citenamefont {Dudek}, \citenamefont {G\'o\'zd\'z},\ and\
		\citenamefont {Regan}}]{Schunck2004_PRC69-061305R}%
	\BibitemOpen
	\bibfield  {author} {\bibinfo {author} {\bibfnamefont {N.}~\bibnamefont
			{Schunck}}, \bibinfo {author} {\bibfnamefont {J.}~\bibnamefont {Dudek}},
		\bibinfo {author} {\bibfnamefont {A.}~\bibnamefont {G\'o\'zd\'z}}, \ and\
		\bibinfo {author} {\bibfnamefont {P.~H.}\ \bibnamefont {Regan}},\ }\href
	{http://link.aps.org/doi/10.1103/PhysRevC.69.061305} {\bibfield  {journal}
		{\bibinfo  {journal} {Phys. Rev. C}\ }\textbf {\bibinfo {volume} {69}},\
		\bibinfo {pages} {061305(R)} (\bibinfo {year} {2004})}\BibitemShut {NoStop}%
	\bibitem [{\citenamefont {Dudek}\ \emph {et~al.}(2014)\citenamefont {Dudek},
		\citenamefont {Curien}, \citenamefont {Rouvel}, \citenamefont {Mazurek},
		\citenamefont {Shimizu},\ and\ \citenamefont
		{Tagami}}]{Dudek2014_PS89-054007}%
	\BibitemOpen
	\bibfield  {author} {\bibinfo {author} {\bibfnamefont {J.}~\bibnamefont
			{Dudek}}, \bibinfo {author} {\bibfnamefont {D.}~\bibnamefont {Curien}},
		\bibinfo {author} {\bibfnamefont {D.}~\bibnamefont {Rouvel}}, \bibinfo
		{author} {\bibfnamefont {K.}~\bibnamefont {Mazurek}}, \bibinfo {author}
		{\bibfnamefont {Y.~R.}\ \bibnamefont {Shimizu}}, \ and\ \bibinfo {author}
		{\bibfnamefont {S.}~\bibnamefont {Tagami}},\ }\href
	{http://stacks.iop.org/1402-4896/89/i=5/a=054007} {\bibfield  {journal}
		{\bibinfo  {journal} {Phys. Scr.}\ }\textbf {\bibinfo {volume} {89}},\
		\bibinfo {pages} {054007} (\bibinfo {year} {2014})}\BibitemShut {NoStop}%
	\bibitem [{\citenamefont {Jachimowicz}\ \emph {et~al.}(2017)\citenamefont
		{Jachimowicz}, \citenamefont {Kowal},\ and\ \citenamefont
		{Skalski}}]{Jachimowicz2017_PRC95-034329}%
	\BibitemOpen
	\bibfield  {author} {\bibinfo {author} {\bibfnamefont {P.}~\bibnamefont
			{Jachimowicz}}, \bibinfo {author} {\bibfnamefont {M.}~\bibnamefont {Kowal}},
		\ and\ \bibinfo {author} {\bibfnamefont {J.}~\bibnamefont {Skalski}},\ }\href
	{https://link.aps.org/doi/10.1103/PhysRevC.95.034329} {\bibfield  {journal}
		{\bibinfo  {journal} {Phys. Rev. C}\ }\textbf {\bibinfo {volume} {95}},\
		\bibinfo {pages} {034329} (\bibinfo {year} {2017})}\BibitemShut {NoStop}%
	\bibitem [{\citenamefont {Yang}\ \emph
		{et~al.}(2022{\natexlab{a}})\citenamefont {Yang}, \citenamefont {Dudek},
		\citenamefont {Dedes}, \citenamefont {Baran}, \citenamefont {Curien},
		\citenamefont {Gaamouci}, \citenamefont {G{\'{o}}{\'{z}}d{\'{z}}},
		\citenamefont {P{\c{e}}drak}, \citenamefont {Rouvel},\ and\ \citenamefont
		{Wang}}]{Yang2022_PRC106-054314}%
	\BibitemOpen
	\bibfield  {author} {\bibinfo {author} {\bibfnamefont {J.}~\bibnamefont
			{Yang}}, \bibinfo {author} {\bibfnamefont {J.}~\bibnamefont {Dudek}},
		\bibinfo {author} {\bibfnamefont {I.}~\bibnamefont {Dedes}}, \bibinfo
		{author} {\bibfnamefont {A.}~\bibnamefont {Baran}}, \bibinfo {author}
		{\bibfnamefont {D.}~\bibnamefont {Curien}}, \bibinfo {author} {\bibfnamefont
			{A.}~\bibnamefont {Gaamouci}}, \bibinfo {author} {\bibfnamefont
			{A.}~\bibnamefont {G{\'{o}}{\'{z}}d{\'{z}}}}, \bibinfo {author}
		{\bibfnamefont {A.}~\bibnamefont {P{\c{e}}drak}}, \bibinfo {author}
		{\bibfnamefont {D.}~\bibnamefont {Rouvel}}, \ and\ \bibinfo {author}
		{\bibfnamefont {H.~L.}\ \bibnamefont {Wang}},\ }\href {\doibase
		10.1103/physrevc.106.054314} {\bibfield  {journal} {\bibinfo  {journal} {Phys
				Rev C}\ }\textbf {\bibinfo {volume} {106}},\ \bibinfo {pages} {054314}
		(\bibinfo {year} {2022}{\natexlab{a}})}\BibitemShut {NoStop}%
	\bibitem [{\citenamefont {Yang}\ \emph
		{et~al.}(2022{\natexlab{b}})\citenamefont {Yang}, \citenamefont {Dudek},
		\citenamefont {Dedes}, \citenamefont {Baran}, \citenamefont {Curien},
		\citenamefont {Gaamouci}, \citenamefont {G{\'{o}}{\'{z}}d{\'{z}}},
		\citenamefont {P{\c{e}}drak}, \citenamefont {Rouvel}, \citenamefont {Wang},\
		and\ \citenamefont {Burkat}}]{Yang2022_PRC105-034348}%
	\BibitemOpen
	\bibfield  {author} {\bibinfo {author} {\bibfnamefont {J.}~\bibnamefont
			{Yang}}, \bibinfo {author} {\bibfnamefont {J.}~\bibnamefont {Dudek}},
		\bibinfo {author} {\bibfnamefont {I.}~\bibnamefont {Dedes}}, \bibinfo
		{author} {\bibfnamefont {A.}~\bibnamefont {Baran}}, \bibinfo {author}
		{\bibfnamefont {D.}~\bibnamefont {Curien}}, \bibinfo {author} {\bibfnamefont
			{A.}~\bibnamefont {Gaamouci}}, \bibinfo {author} {\bibfnamefont
			{A.}~\bibnamefont {G{\'{o}}{\'{z}}d{\'{z}}}}, \bibinfo {author}
		{\bibfnamefont {A.}~\bibnamefont {P{\c{e}}drak}}, \bibinfo {author}
		{\bibfnamefont {D.}~\bibnamefont {Rouvel}}, \bibinfo {author} {\bibfnamefont
			{H.~L.}\ \bibnamefont {Wang}}, \ and\ \bibinfo {author} {\bibfnamefont
			{J.}~\bibnamefont {Burkat}},\ }\href {\doibase 10.1103/physrevc.105.034348}
	{\bibfield  {journal} {\bibinfo  {journal} {Phys. Rev. C}\ }\textbf {\bibinfo
			{volume} {105}},\ \bibinfo {pages} {034348} (\bibinfo {year}
		{2022}{\natexlab{b}})}\BibitemShut {NoStop}%
	\bibitem [{\citenamefont {Yamagami}\ \emph {et~al.}(2001)\citenamefont
		{Yamagami}, \citenamefont {Matsuyanagi},\ and\ \citenamefont
		{Matsuo}}]{Yamagami2001_NPA693-579}%
	\BibitemOpen
	\bibfield  {author} {\bibinfo {author} {\bibfnamefont {M.}~\bibnamefont
			{Yamagami}}, \bibinfo {author} {\bibfnamefont {K.}~\bibnamefont
			{Matsuyanagi}}, \ and\ \bibinfo {author} {\bibfnamefont {M.}~\bibnamefont
			{Matsuo}},\ }\href
	{http://www.sciencedirect.com/science/article/pii/S0375947401009186}
	{\bibfield  {journal} {\bibinfo  {journal} {Nucl. Phys. A}\ }\textbf
		{\bibinfo {volume} {693}},\ \bibinfo {pages} {579} (\bibinfo {year}
		{2001})}\BibitemShut {NoStop}%
	\bibitem [{\citenamefont {Olbratowski}\ \emph {et~al.}(2006)\citenamefont
		{Olbratowski}, \citenamefont {Dobaczewski}, \citenamefont {Powalowski},
		\citenamefont {Sadziak},\ and\ \citenamefont
		{Zberecki}}]{Olbratowski2006_IJMPE15-333}%
	\BibitemOpen
	\bibfield  {author} {\bibinfo {author} {\bibfnamefont {P.}~\bibnamefont
			{Olbratowski}}, \bibinfo {author} {\bibfnamefont {J.}~\bibnamefont
			{Dobaczewski}}, \bibinfo {author} {\bibfnamefont {P.}~\bibnamefont
			{Powalowski}}, \bibinfo {author} {\bibfnamefont {M.}~\bibnamefont {Sadziak}},
		\ and\ \bibinfo {author} {\bibfnamefont {K.}~\bibnamefont {Zberecki}},\
	}\href {\doibase 10.1142/S021830130600417X} {\bibfield  {journal} {\bibinfo
			{journal} {Int. J. Mod. Phys. E}\ }\textbf {\bibinfo {volume} {15}},\
		\bibinfo {pages} {333} (\bibinfo {year} {2006})}\BibitemShut {NoStop}%
	\bibitem [{\citenamefont {Zberecki}\ \emph {et~al.}(2009)\citenamefont
		{Zberecki}, \citenamefont {Heenen},\ and\ \citenamefont
		{Magierski}}]{Zberecki2009_PRC79-014319}%
	\BibitemOpen
	\bibfield  {author} {\bibinfo {author} {\bibfnamefont {K.}~\bibnamefont
			{Zberecki}}, \bibinfo {author} {\bibfnamefont {P.-H.}\ \bibnamefont
			{Heenen}}, \ and\ \bibinfo {author} {\bibfnamefont {P.}~\bibnamefont
			{Magierski}},\ }\href {http://link.aps.org/doi/10.1103/PhysRevC.79.014319}
	{\bibfield  {journal} {\bibinfo  {journal} {Phys. Rev. C}\ }\textbf {\bibinfo
			{volume} {79}},\ \bibinfo {pages} {014319} (\bibinfo {year}
		{2009})}\BibitemShut {NoStop}%
	\bibitem [{\citenamefont {Zberecki}\ \emph {et~al.}(2006)\citenamefont
		{Zberecki}, \citenamefont {Magierski}, \citenamefont {Heenen},\ and\
		\citenamefont {Schunck}}]{Zberecki2006_PRC74-051302R}%
	\BibitemOpen
	\bibfield  {author} {\bibinfo {author} {\bibfnamefont {K.}~\bibnamefont
			{Zberecki}}, \bibinfo {author} {\bibfnamefont {P.}~\bibnamefont {Magierski}},
		\bibinfo {author} {\bibfnamefont {P.-H.}\ \bibnamefont {Heenen}}, \ and\
		\bibinfo {author} {\bibfnamefont {N.}~\bibnamefont {Schunck}},\ }\href
	{http://link.aps.org/doi/10.1103/PhysRevC.74.051302} {\bibfield  {journal}
		{\bibinfo  {journal} {Phys. Rev. C}\ }\textbf {\bibinfo {volume} {74}},\
		\bibinfo {pages} {051302(R)} (\bibinfo {year} {2006})}\BibitemShut {NoStop}%
	\bibitem [{\citenamefont {Takami}\ \emph {et~al.}(1998)\citenamefont {Takami},
		\citenamefont {Yabana},\ and\ \citenamefont
		{Matsuo}}]{Takami1998_PLB431-242}%
	\BibitemOpen
	\bibfield  {author} {\bibinfo {author} {\bibfnamefont {S.}~\bibnamefont
			{Takami}}, \bibinfo {author} {\bibfnamefont {K.}~\bibnamefont {Yabana}}, \
		and\ \bibinfo {author} {\bibfnamefont {M.}~\bibnamefont {Matsuo}},\ }\href
	{http://www.sciencedirect.com/science/article/pii/S0370269398005450}
	{\bibfield  {journal} {\bibinfo  {journal} {Phys. Lett. B}\ }\textbf
		{\bibinfo {volume} {431}},\ \bibinfo {pages} {242} (\bibinfo {year}
		{1998})}\BibitemShut {NoStop}%
	\bibitem [{\citenamefont {Wang}\ \emph {et~al.}(2019)\citenamefont {Wang},
		\citenamefont {Dong}, \citenamefont {Gao}, \citenamefont {Chen},\ and\
		\citenamefont {Shen}}]{Wang2019_PLB790-498}%
	\BibitemOpen
	\bibfield  {author} {\bibinfo {author} {\bibfnamefont {X.~B.}\ \bibnamefont
			{Wang}}, \bibinfo {author} {\bibfnamefont {G.~X.}\ \bibnamefont {Dong}},
		\bibinfo {author} {\bibfnamefont {Z.~C.}\ \bibnamefont {Gao}}, \bibinfo
		{author} {\bibfnamefont {Y.~S.}\ \bibnamefont {Chen}}, \ and\ \bibinfo
		{author} {\bibfnamefont {C.~W.}\ \bibnamefont {Shen}},\ }\href
	{http://www.sciencedirect.com/science/article/pii/S0370269319300905}
	{\bibfield  {journal} {\bibinfo  {journal} {Phys. Lett. B}\ }\textbf
		{\bibinfo {volume} {790}},\ \bibinfo {pages} {498} (\bibinfo {year}
		{2019})}\BibitemShut {NoStop}%
	\bibitem [{\citenamefont {Zhao}\ \emph {et~al.}(2017)\citenamefont {Zhao},
		\citenamefont {Lu}, \citenamefont {Zhao},\ and\ \citenamefont
		{Zhou}}]{Zhao2017_PRC95-014320}%
	\BibitemOpen
	\bibfield  {author} {\bibinfo {author} {\bibfnamefont {J.}~\bibnamefont
			{Zhao}}, \bibinfo {author} {\bibfnamefont {B.-N.}\ \bibnamefont {Lu}},
		\bibinfo {author} {\bibfnamefont {E.-G.}\ \bibnamefont {Zhao}}, \ and\
		\bibinfo {author} {\bibfnamefont {S.-G.}\ \bibnamefont {Zhou}},\ }\href
	{\doibase 10.1103/PhysRevC.95.014320} {\bibfield  {journal} {\bibinfo
			{journal} {Phys. Rev. C}\ }\textbf {\bibinfo {volume} {95}},\ \bibinfo
		{pages} {014320} (\bibinfo {year} {2017})}\BibitemShut {NoStop}%
	\bibitem [{\citenamefont {Zhao}\ and\ \citenamefont
		{Wu}(2024)}]{Zhao2024_PRC109-014303}%
	\BibitemOpen
	\bibfield  {author} {\bibinfo {author} {\bibfnamefont {J.}~\bibnamefont
			{Zhao}}\ and\ \bibinfo {author} {\bibfnamefont {Z.-G.}\ \bibnamefont {Wu}},\
	}\href {\doibase 10.1103/physrevc.109.014303} {\bibfield  {journal} {\bibinfo
			{journal} {Phys. Rev. C}\ }\textbf {\bibinfo {volume} {109}},\ \bibinfo
		{pages} {014303} (\bibinfo {year} {2024})}\BibitemShut {NoStop}%
	\bibitem [{\citenamefont {Xu}\ \emph {et~al.}(2024)\citenamefont {Xu},
		\citenamefont {Li}, \citenamefont {Ren},\ and\ \citenamefont
		{Zhao}}]{Xu2024_PRC109-014311}%
	\BibitemOpen
	\bibfield  {author} {\bibinfo {author} {\bibfnamefont {F.~F.}\ \bibnamefont
			{Xu}}, \bibinfo {author} {\bibfnamefont {B.}~\bibnamefont {Li}}, \bibinfo
		{author} {\bibfnamefont {Z.~X.}\ \bibnamefont {Ren}}, \ and\ \bibinfo
		{author} {\bibfnamefont {P.~W.}\ \bibnamefont {Zhao}},\ }\href {\doibase
		10.1103/physrevc.109.014311} {\bibfield  {journal} {\bibinfo  {journal}
			{Phys. Rev. C}\ }\textbf {\bibinfo {volume} {109}},\ \bibinfo {pages}
		{014311} (\bibinfo {year} {2024})}\BibitemShut {NoStop}%
	\bibitem [{\citenamefont {Bijker}\ and\ \citenamefont
		{Iachello}(2014)}]{Bijker2014_PRL112-152501}%
	\BibitemOpen
	\bibfield  {author} {\bibinfo {author} {\bibfnamefont {R.}~\bibnamefont
			{Bijker}}\ and\ \bibinfo {author} {\bibfnamefont {F.}~\bibnamefont
			{Iachello}},\ }\href {http://link.aps.org/doi/10.1103/PhysRevLett.112.152501}
	{\bibfield  {journal} {\bibinfo  {journal} {Phys. Rev. Lett.}\ }\textbf
		{\bibinfo {volume} {112}},\ \bibinfo {pages} {152501} (\bibinfo {year}
		{2014})}\BibitemShut {NoStop}%
	\bibitem [{\citenamefont {Gao}\ \emph {et~al.}(2004)\citenamefont {Gao},
		\citenamefont {Chen},\ and\ \citenamefont {Meng}}]{Gao2004_CPL21-806}%
	\BibitemOpen
	\bibfield  {author} {\bibinfo {author} {\bibfnamefont {Z.-C.}\ \bibnamefont
			{Gao}}, \bibinfo {author} {\bibfnamefont {Y.-S.}\ \bibnamefont {Chen}}, \
		and\ \bibinfo {author} {\bibfnamefont {J.}~\bibnamefont {Meng}},\ }\href
	{http://stacks.iop.org/0256-307X/21/i=5/a=012} {\bibfield  {journal}
		{\bibinfo  {journal} {Chin. Phys. Lett.}\ }\textbf {\bibinfo {volume} {21}},\
		\bibinfo {pages} {806} (\bibinfo {year} {2004})}\BibitemShut {NoStop}%
	\bibitem [{\citenamefont {Chen}\ and\ \citenamefont
		{Gao}(2010)}]{Chen2010_NPA834-378c}%
	\BibitemOpen
	\bibfield  {author} {\bibinfo {author} {\bibfnamefont {Y.}~\bibnamefont
			{Chen}}\ and\ \bibinfo {author} {\bibfnamefont {Z.-C.}\ \bibnamefont {Gao}},\
	}\bibfield  {booktitle} {\emph {\bibinfo {booktitle} {The 10th International
				Conference on Nucleus-Nucleus Collisions (NN2009)}},\ }\href
	{http://www.sciencedirect.com/science/article/pii/S037594741000045X}
	{\bibfield  {journal} {\bibinfo  {journal} {Nucl. Phys. A}\ }\textbf
		{\bibinfo {volume} {834}},\ \bibinfo {pages} {378c} (\bibinfo {year}
		{2010})}\BibitemShut {NoStop}%
	\bibitem [{\citenamefont {Epelbaum}\ \emph {et~al.}(2014)\citenamefont
		{Epelbaum}, \citenamefont {Krebs}, \citenamefont {L\"ahde}, \citenamefont
		{Lee}, \citenamefont {Mei{\ss}ner},\ and\ \citenamefont
		{Rupak}}]{Epelbaum2014_PRL112-102501}%
	\BibitemOpen
	\bibfield  {author} {\bibinfo {author} {\bibfnamefont {E.}~\bibnamefont
			{Epelbaum}}, \bibinfo {author} {\bibfnamefont {H.}~\bibnamefont {Krebs}},
		\bibinfo {author} {\bibfnamefont {T.~A.}\ \bibnamefont {L\"ahde}}, \bibinfo
		{author} {\bibfnamefont {D.}~\bibnamefont {Lee}}, \bibinfo {author}
		{\bibfnamefont {U.-G.}\ \bibnamefont {Mei{\ss}ner}}, \ and\ \bibinfo {author}
		{\bibfnamefont {G.}~\bibnamefont {Rupak}},\ }\href
	{http://link.aps.org/doi/10.1103/PhysRevLett.112.102501} {\bibfield
		{journal} {\bibinfo  {journal} {Phys. Rev. Lett.}\ }\textbf {\bibinfo
			{volume} {112}},\ \bibinfo {pages} {102501} (\bibinfo {year}
		{2014})}\BibitemShut {NoStop}%
	\bibitem [{\citenamefont {Tagami}\ \emph {et~al.}(2013)\citenamefont {Tagami},
		\citenamefont {Shimizu},\ and\ \citenamefont
		{Dudek}}]{Tagami2013_PRC87-054306}%
	\BibitemOpen
	\bibfield  {author} {\bibinfo {author} {\bibfnamefont {S.}~\bibnamefont
			{Tagami}}, \bibinfo {author} {\bibfnamefont {Y.~R.}\ \bibnamefont {Shimizu}},
		\ and\ \bibinfo {author} {\bibfnamefont {J.}~\bibnamefont {Dudek}},\ }\href
	{http://link.aps.org/doi/10.1103/PhysRevC.87.054306} {\bibfield  {journal}
		{\bibinfo  {journal} {Phys. Rev. C}\ }\textbf {\bibinfo {volume} {87}},\
		\bibinfo {pages} {054306} (\bibinfo {year} {2013})}\BibitemShut {NoStop}%
	\bibitem [{\citenamefont {Tagami}\ \emph {et~al.}(2015)\citenamefont {Tagami},
		\citenamefont {Shimizu},\ and\ \citenamefont
		{Dudek}}]{Tagami2015_JPG42-015106}%
	\BibitemOpen
	\bibfield  {author} {\bibinfo {author} {\bibfnamefont {S.}~\bibnamefont
			{Tagami}}, \bibinfo {author} {\bibfnamefont {Y.~R.}\ \bibnamefont {Shimizu}},
		\ and\ \bibinfo {author} {\bibfnamefont {J.}~\bibnamefont {Dudek}},\ }\href
	{http://stacks.iop.org/0954-3899/42/i=1/a=015106} {\bibfield  {journal}
		{\bibinfo  {journal} {J. Phys. G: Nucl. Part. Phys.}\ }\textbf {\bibinfo
			{volume} {42}},\ \bibinfo {pages} {015106} (\bibinfo {year}
		{2015})}\BibitemShut {NoStop}%
	\bibitem [{\citenamefont {Chen}\ and\ \citenamefont
		{Gao}(2013)}]{Chen2013_NPR30-278}%
	\BibitemOpen
	\bibfield  {author} {\bibinfo {author} {\bibfnamefont {Y.}~\bibnamefont
			{Chen}}\ and\ \bibinfo {author} {\bibfnamefont {Z.}~\bibnamefont {Gao}},\
	}\href {\doibase 10.11804/NuclPhysRev.30.03.278} {\bibfield  {journal}
		{\bibinfo  {journal} {Nucl. Phys. Rev.}\ }\textbf {\bibinfo {volume} {30}},\
		\bibinfo {pages} {278} (\bibinfo {year} {2013})}\BibitemShut {NoStop}%
	\bibitem [{\citenamefont {Tagami}\ \emph {et~al.}(2018)\citenamefont {Tagami},
		\citenamefont {Shimizu},\ and\ \citenamefont
		{Dudek}}]{Tagami2018_PRC98-024304}%
	\BibitemOpen
	\bibfield  {author} {\bibinfo {author} {\bibfnamefont {S.}~\bibnamefont
			{Tagami}}, \bibinfo {author} {\bibfnamefont {Y.~R.}\ \bibnamefont {Shimizu}},
		\ and\ \bibinfo {author} {\bibfnamefont {J.}~\bibnamefont {Dudek}},\ }\href
	{https://link.aps.org/doi/10.1103/PhysRevC.98.024304} {\bibfield  {journal}
		{\bibinfo  {journal} {Phys. Rev. C}\ }\textbf {\bibinfo {volume} {98}},\
		\bibinfo {pages} {024304} (\bibinfo {year} {2018})}\BibitemShut {NoStop}%
	\bibitem [{\citenamefont {Bark}\ \emph {et~al.}(2010)\citenamefont {Bark},
		\citenamefont {Sharpey-Schafer}, \citenamefont {Maliage}, \citenamefont
		{Madiba}, \citenamefont {Komati}, \citenamefont {Lawrie}, \citenamefont
		{Lawrie}, \citenamefont {Lindsay}, \citenamefont {Maine}, \citenamefont
		{Mullins}, \citenamefont {Murray}, \citenamefont {Ncapayi}, \citenamefont
		{Ramashidza}, \citenamefont {Smit},\ and\ \citenamefont
		{Vymers}}]{Bark2010_PRL104-022501}%
	\BibitemOpen
	\bibfield  {author} {\bibinfo {author} {\bibfnamefont {R.~A.}\ \bibnamefont
			{Bark}}, \bibinfo {author} {\bibfnamefont {J.~F.}\ \bibnamefont
			{Sharpey-Schafer}}, \bibinfo {author} {\bibfnamefont {S.~M.}\ \bibnamefont
			{Maliage}}, \bibinfo {author} {\bibfnamefont {T.~E.}\ \bibnamefont {Madiba}},
		\bibinfo {author} {\bibfnamefont {F.~S.}\ \bibnamefont {Komati}}, \bibinfo
		{author} {\bibfnamefont {E.~A.}\ \bibnamefont {Lawrie}}, \bibinfo {author}
		{\bibfnamefont {J.~J.}\ \bibnamefont {Lawrie}}, \bibinfo {author}
		{\bibfnamefont {R.}~\bibnamefont {Lindsay}}, \bibinfo {author} {\bibfnamefont
			{P.}~\bibnamefont {Maine}}, \bibinfo {author} {\bibfnamefont {S.~M.}\
			\bibnamefont {Mullins}}, \bibinfo {author} {\bibfnamefont {S.~H.~T.}\
			\bibnamefont {Murray}}, \bibinfo {author} {\bibfnamefont {N.~J.}\
			\bibnamefont {Ncapayi}}, \bibinfo {author} {\bibfnamefont {T.~M.}\
			\bibnamefont {Ramashidza}}, \bibinfo {author} {\bibfnamefont {F.~D.}\
			\bibnamefont {Smit}}, \ and\ \bibinfo {author} {\bibfnamefont
			{P.}~\bibnamefont {Vymers}},\ }\href
	{http://link.aps.org/doi/10.1103/PhysRevLett.104.022501} {\bibfield
		{journal} {\bibinfo  {journal} {Phys. Rev. Lett.}\ }\textbf {\bibinfo
			{volume} {104}},\ \bibinfo {pages} {022501} (\bibinfo {year}
		{2010})}\BibitemShut {NoStop}%
	\bibitem [{\citenamefont {Jentschel}\ \emph {et~al.}(2010)\citenamefont
		{Jentschel}, \citenamefont {Urban}, \citenamefont {Krempel}, \citenamefont
		{Tonev}, \citenamefont {Dudek}, \citenamefont {Curien}, \citenamefont
		{Lauss}, \citenamefont {de~Angelis},\ and\ \citenamefont
		{Petkov}}]{Jentschel2010_PRL104-222502}%
	\BibitemOpen
	\bibfield  {author} {\bibinfo {author} {\bibfnamefont {M.}~\bibnamefont
			{Jentschel}}, \bibinfo {author} {\bibfnamefont {W.}~\bibnamefont {Urban}},
		\bibinfo {author} {\bibfnamefont {J.}~\bibnamefont {Krempel}}, \bibinfo
		{author} {\bibfnamefont {D.}~\bibnamefont {Tonev}}, \bibinfo {author}
		{\bibfnamefont {J.}~\bibnamefont {Dudek}}, \bibinfo {author} {\bibfnamefont
			{D.}~\bibnamefont {Curien}}, \bibinfo {author} {\bibfnamefont
			{B.}~\bibnamefont {Lauss}}, \bibinfo {author} {\bibfnamefont
			{G.}~\bibnamefont {de~Angelis}}, \ and\ \bibinfo {author} {\bibfnamefont
			{P.}~\bibnamefont {Petkov}},\ }\href
	{http://link.aps.org/doi/10.1103/PhysRevLett.104.222502} {\bibfield
		{journal} {\bibinfo  {journal} {Phys. Rev. Lett.}\ }\textbf {\bibinfo
			{volume} {104}},\ \bibinfo {pages} {222502} (\bibinfo {year}
		{2010})}\BibitemShut {NoStop}%
	\bibitem [{\citenamefont {Doan}\ \emph {et~al.}(2010)\citenamefont {Doan},
		\citenamefont {Vancraeyenest}, \citenamefont {Stezowski}, \citenamefont
		{Guinet}, \citenamefont {Curien}, \citenamefont {Dudek}, \citenamefont
		{Lautesse}, \citenamefont {Lehaut}, \citenamefont {Redon}, \citenamefont
		{Schmitt}, \citenamefont {Duchene}, \citenamefont {Gall}, \citenamefont
		{Molique}, \citenamefont {Piot}, \citenamefont {Greenlees}, \citenamefont
		{Jakobsson}, \citenamefont {Julin}, \citenamefont {Juutinen}, \citenamefont
		{Jones}, \citenamefont {Ketelhut}, \citenamefont {Nyman}, \citenamefont
		{Peura}, \citenamefont {Rahkila}, \citenamefont {Gozdz}, \citenamefont
		{Mazurek}, \citenamefont {Schunck}, \citenamefont {Zuber}, \citenamefont
		{Bednarczyk}, \citenamefont {Maj}, \citenamefont {Astier}, \citenamefont
		{Deloncle}, \citenamefont {Verney}, \citenamefont {de~Angelis},\ and\
		\citenamefont {Gerl}}]{Doan2010_PRC82-067306}%
	\BibitemOpen
	\bibfield  {author} {\bibinfo {author} {\bibfnamefont {Q.~T.}\ \bibnamefont
			{Doan}}, \bibinfo {author} {\bibfnamefont {A.}~\bibnamefont {Vancraeyenest}},
		\bibinfo {author} {\bibfnamefont {O.}~\bibnamefont {Stezowski}}, \bibinfo
		{author} {\bibfnamefont {D.}~\bibnamefont {Guinet}}, \bibinfo {author}
		{\bibfnamefont {D.}~\bibnamefont {Curien}}, \bibinfo {author} {\bibfnamefont
			{J.}~\bibnamefont {Dudek}}, \bibinfo {author} {\bibfnamefont
			{P.}~\bibnamefont {Lautesse}}, \bibinfo {author} {\bibfnamefont
			{G.}~\bibnamefont {Lehaut}}, \bibinfo {author} {\bibfnamefont
			{N.}~\bibnamefont {Redon}}, \bibinfo {author} {\bibfnamefont
			{C.}~\bibnamefont {Schmitt}}, \bibinfo {author} {\bibfnamefont
			{G.}~\bibnamefont {Duchene}}, \bibinfo {author} {\bibfnamefont
			{B.}~\bibnamefont {Gall}}, \bibinfo {author} {\bibfnamefont {H.}~\bibnamefont
			{Molique}}, \bibinfo {author} {\bibfnamefont {J.}~\bibnamefont {Piot}},
		\bibinfo {author} {\bibfnamefont {P.~T.}\ \bibnamefont {Greenlees}}, \bibinfo
		{author} {\bibfnamefont {U.}~\bibnamefont {Jakobsson}}, \bibinfo {author}
		{\bibfnamefont {R.}~\bibnamefont {Julin}}, \bibinfo {author} {\bibfnamefont
			{S.}~\bibnamefont {Juutinen}}, \bibinfo {author} {\bibfnamefont
			{P.}~\bibnamefont {Jones}}, \bibinfo {author} {\bibfnamefont
			{S.}~\bibnamefont {Ketelhut}}, \bibinfo {author} {\bibfnamefont
			{M.}~\bibnamefont {Nyman}}, \bibinfo {author} {\bibfnamefont
			{P.}~\bibnamefont {Peura}}, \bibinfo {author} {\bibfnamefont
			{P.}~\bibnamefont {Rahkila}}, \bibinfo {author} {\bibfnamefont
			{A.}~\bibnamefont {Gozdz}}, \bibinfo {author} {\bibfnamefont
			{K.}~\bibnamefont {Mazurek}}, \bibinfo {author} {\bibfnamefont
			{N.}~\bibnamefont {Schunck}}, \bibinfo {author} {\bibfnamefont
			{K.}~\bibnamefont {Zuber}}, \bibinfo {author} {\bibfnamefont
			{P.}~\bibnamefont {Bednarczyk}}, \bibinfo {author} {\bibfnamefont
			{A.}~\bibnamefont {Maj}}, \bibinfo {author} {\bibfnamefont {A.}~\bibnamefont
			{Astier}}, \bibinfo {author} {\bibfnamefont {I.}~\bibnamefont {Deloncle}},
		\bibinfo {author} {\bibfnamefont {D.}~\bibnamefont {Verney}}, \bibinfo
		{author} {\bibfnamefont {G.}~\bibnamefont {de~Angelis}}, \ and\ \bibinfo
		{author} {\bibfnamefont {J.}~\bibnamefont {Gerl}},\ }\href
	{http://link.aps.org/doi/10.1103/PhysRevC.82.067306} {\bibfield  {journal}
		{\bibinfo  {journal} {Phys. Rev. C}\ }\textbf {\bibinfo {volume} {82}},\
		\bibinfo {pages} {067306} (\bibinfo {year} {2010})}\BibitemShut {NoStop}%
	\bibitem [{\citenamefont {Ntshangase}\ \emph {et~al.}(2010)\citenamefont
		{Ntshangase}, \citenamefont {Bark}, \citenamefont {Aschman}, \citenamefont
		{Bvumbi}, \citenamefont {Datta}, \citenamefont {Davidson}, \citenamefont
		{Dinoko}, \citenamefont {Elbasher}, \citenamefont {Juhasz}, \citenamefont
		{Khaleel}, \citenamefont {Krasznahorkay}, \citenamefont {Lawrie},
		\citenamefont {Lawrie}, \citenamefont {Lieder}, \citenamefont {Majola},
		\citenamefont {Masiteng}, \citenamefont {Mohammed}, \citenamefont {Mullins},
		\citenamefont {Nieminen}, \citenamefont {Nyako}, \citenamefont {Papka},
		\citenamefont {Roux}, \citenamefont {Sharpey-Shafer}, \citenamefont
		{Shirinda}, \citenamefont {Stankiewicz}, \citenamefont {Timar},\ and\
		\citenamefont {Wilson}}]{Ntshangase2010_PRC82-041305R}%
	\BibitemOpen
	\bibfield  {author} {\bibinfo {author} {\bibfnamefont {S.~S.}\ \bibnamefont
			{Ntshangase}}, \bibinfo {author} {\bibfnamefont {R.~A.}\ \bibnamefont
			{Bark}}, \bibinfo {author} {\bibfnamefont {D.~G.}\ \bibnamefont {Aschman}},
		\bibinfo {author} {\bibfnamefont {S.}~\bibnamefont {Bvumbi}}, \bibinfo
		{author} {\bibfnamefont {P.}~\bibnamefont {Datta}}, \bibinfo {author}
		{\bibfnamefont {P.~M.}\ \bibnamefont {Davidson}}, \bibinfo {author}
		{\bibfnamefont {T.~S.}\ \bibnamefont {Dinoko}}, \bibinfo {author}
		{\bibfnamefont {M.~E.~A.}\ \bibnamefont {Elbasher}}, \bibinfo {author}
		{\bibfnamefont {K.}~\bibnamefont {Juhasz}}, \bibinfo {author} {\bibfnamefont
			{E.~M.~A.}\ \bibnamefont {Khaleel}}, \bibinfo {author} {\bibfnamefont
			{A.}~\bibnamefont {Krasznahorkay}}, \bibinfo {author} {\bibfnamefont {E.~A.}\
			\bibnamefont {Lawrie}}, \bibinfo {author} {\bibfnamefont {J.~J.}\
			\bibnamefont {Lawrie}}, \bibinfo {author} {\bibfnamefont {R.~M.}\
			\bibnamefont {Lieder}}, \bibinfo {author} {\bibfnamefont {S.~N.~T.}\
			\bibnamefont {Majola}}, \bibinfo {author} {\bibfnamefont {P.~L.}\
			\bibnamefont {Masiteng}}, \bibinfo {author} {\bibfnamefont {H.}~\bibnamefont
			{Mohammed}}, \bibinfo {author} {\bibfnamefont {S.~M.}\ \bibnamefont
			{Mullins}}, \bibinfo {author} {\bibfnamefont {P.}~\bibnamefont {Nieminen}},
		\bibinfo {author} {\bibfnamefont {B.~M.}\ \bibnamefont {Nyako}}, \bibinfo
		{author} {\bibfnamefont {P.}~\bibnamefont {Papka}}, \bibinfo {author}
		{\bibfnamefont {D.~G.}\ \bibnamefont {Roux}}, \bibinfo {author}
		{\bibfnamefont {J.~F.}\ \bibnamefont {Sharpey-Shafer}}, \bibinfo {author}
		{\bibfnamefont {O.}~\bibnamefont {Shirinda}}, \bibinfo {author}
		{\bibfnamefont {M.~A.}\ \bibnamefont {Stankiewicz}}, \bibinfo {author}
		{\bibfnamefont {J.}~\bibnamefont {Timar}}, \ and\ \bibinfo {author}
		{\bibfnamefont {A.~N.}\ \bibnamefont {Wilson}},\ }\href
	{http://link.aps.org/doi/10.1103/PhysRevC.82.041305} {\bibfield  {journal}
		{\bibinfo  {journal} {Phys. Rev. C}\ }\textbf {\bibinfo {volume} {82}},\
		\bibinfo {pages} {041305(R)} (\bibinfo {year} {2010})}\BibitemShut {NoStop}%
	\bibitem [{\citenamefont {Hartley}\ \emph {et~al.}(2017)\citenamefont
		{Hartley}, \citenamefont {Riedinger}, \citenamefont {Janssens}, \citenamefont
		{Majola}, \citenamefont {Riley}, \citenamefont {Allmond}, \citenamefont
		{Beausang}, \citenamefont {Carpenter}, \citenamefont {Chiara}, \citenamefont
		{Cooper}, \citenamefont {Curien}, \citenamefont {Gall}, \citenamefont
		{Garrett}, \citenamefont {Kondev}, \citenamefont {Kulp}, \citenamefont
		{Lauritsen}, \citenamefont {McCutchan}, \citenamefont {Miller}, \citenamefont
		{Miller}, \citenamefont {Piot}, \citenamefont {Redon}, \citenamefont
		{Sharpey-Schafer}, \citenamefont {Simpson}, \citenamefont {Stefanescu},
		\citenamefont {Wang}, \citenamefont {Werner}, \citenamefont {Wood},
		\citenamefont {Yu}, \citenamefont {Zhu},\ and\ \citenamefont
		{Dudek}}]{Hartley2017_PRC95-014321}%
	\BibitemOpen
	\bibfield  {author} {\bibinfo {author} {\bibfnamefont {D.~J.}\ \bibnamefont
			{Hartley}}, \bibinfo {author} {\bibfnamefont {L.~L.}\ \bibnamefont
			{Riedinger}}, \bibinfo {author} {\bibfnamefont {R.~V.~F.}\ \bibnamefont
			{Janssens}}, \bibinfo {author} {\bibfnamefont {S.~N.~T.}\ \bibnamefont
			{Majola}}, \bibinfo {author} {\bibfnamefont {M.~A.}\ \bibnamefont {Riley}},
		\bibinfo {author} {\bibfnamefont {J.~M.}\ \bibnamefont {Allmond}}, \bibinfo
		{author} {\bibfnamefont {C.~W.}\ \bibnamefont {Beausang}}, \bibinfo {author}
		{\bibfnamefont {M.~P.}\ \bibnamefont {Carpenter}}, \bibinfo {author}
		{\bibfnamefont {C.~J.}\ \bibnamefont {Chiara}}, \bibinfo {author}
		{\bibfnamefont {N.}~\bibnamefont {Cooper}}, \bibinfo {author} {\bibfnamefont
			{D.}~\bibnamefont {Curien}}, \bibinfo {author} {\bibfnamefont {B.~J.~P.}\
			\bibnamefont {Gall}}, \bibinfo {author} {\bibfnamefont {P.~E.}\ \bibnamefont
			{Garrett}}, \bibinfo {author} {\bibfnamefont {F.~G.}\ \bibnamefont {Kondev}},
		\bibinfo {author} {\bibfnamefont {W.~D.}\ \bibnamefont {Kulp}}, \bibinfo
		{author} {\bibfnamefont {T.}~\bibnamefont {Lauritsen}}, \bibinfo {author}
		{\bibfnamefont {E.~A.}\ \bibnamefont {McCutchan}}, \bibinfo {author}
		{\bibfnamefont {D.}~\bibnamefont {Miller}}, \bibinfo {author} {\bibfnamefont
			{S.}~\bibnamefont {Miller}}, \bibinfo {author} {\bibfnamefont
			{J.}~\bibnamefont {Piot}}, \bibinfo {author} {\bibfnamefont {N.}~\bibnamefont
			{Redon}}, \bibinfo {author} {\bibfnamefont {J.~F.}\ \bibnamefont
			{Sharpey-Schafer}}, \bibinfo {author} {\bibfnamefont {J.}~\bibnamefont
			{Simpson}}, \bibinfo {author} {\bibfnamefont {I.}~\bibnamefont {Stefanescu}},
		\bibinfo {author} {\bibfnamefont {X.}~\bibnamefont {Wang}}, \bibinfo {author}
		{\bibfnamefont {V.}~\bibnamefont {Werner}}, \bibinfo {author} {\bibfnamefont
			{J.~L.}\ \bibnamefont {Wood}}, \bibinfo {author} {\bibfnamefont {C.-H.}\
			\bibnamefont {Yu}}, \bibinfo {author} {\bibfnamefont {S.}~\bibnamefont
			{Zhu}}, \ and\ \bibinfo {author} {\bibfnamefont {J.}~\bibnamefont {Dudek}},\
	}\href {\doibase 10.1103/physrevc.95.014321} {\bibfield  {journal} {\bibinfo
			{journal} {Phys. Rev. C}\ }\textbf {\bibinfo {volume} {95}},\ \bibinfo
		{pages} {014321} (\bibinfo {year} {2017})}\BibitemShut {NoStop}%
	\bibitem [{\citenamefont {Dobrowolski}\ \emph {et~al.}(2011)\citenamefont
		{Dobrowolski}, \citenamefont {G{\'o}{\'z}d{\'z}}, \citenamefont {Mazurek},\
		and\ \citenamefont {Dudek}}]{Dobrowolski2011_IJMPE20-500}%
	\BibitemOpen
	\bibfield  {author} {\bibinfo {author} {\bibfnamefont {A.}~\bibnamefont
			{Dobrowolski}}, \bibinfo {author} {\bibfnamefont {A.}~\bibnamefont
			{G{\'o}{\'z}d{\'z}}}, \bibinfo {author} {\bibfnamefont {K.}~\bibnamefont
			{Mazurek}}, \ and\ \bibinfo {author} {\bibfnamefont {J.}~\bibnamefont
			{Dudek}},\ }\href {\doibase 10.1142/s0218301311017910} {\bibfield  {journal}
		{\bibinfo  {journal} {Int. J. Mod. Phys. E}\ }\textbf {\bibinfo {volume}
			{20}},\ \bibinfo {pages} {500} (\bibinfo {year} {2011})}\BibitemShut
	{NoStop}%
	\bibitem [{\citenamefont {Garg}\ and\ \citenamefont
		{Col{\`o}}(2018)}]{Garg2018_PPNP101-55}%
	\BibitemOpen
	\bibfield  {author} {\bibinfo {author} {\bibfnamefont {U.}~\bibnamefont
			{Garg}}\ and\ \bibinfo {author} {\bibfnamefont {G.}~\bibnamefont
			{Col{\`o}}},\ }\href {\doibase 10.1016/j.ppnp.2018.03.001} {\bibfield
		{journal} {\bibinfo  {journal} {Prog. Part. Nucl. Phys.}\ }\textbf {\bibinfo
			{volume} {101}},\ \bibinfo {pages} {55} (\bibinfo {year} {2018})}\BibitemShut
	{NoStop}%
	\bibitem [{\citenamefont {Garg}\ \emph {et~al.}(1980)\citenamefont {Garg},
		\citenamefont {Bogucki}, \citenamefont {Bronson}, \citenamefont {Lui},
		\citenamefont {Rozsa},\ and\ \citenamefont
		{Youngblood}}]{Garg1980_PRL45-1670}%
	\BibitemOpen
	\bibfield  {author} {\bibinfo {author} {\bibfnamefont {U.}~\bibnamefont
			{Garg}}, \bibinfo {author} {\bibfnamefont {P.}~\bibnamefont {Bogucki}},
		\bibinfo {author} {\bibfnamefont {J.~D.}\ \bibnamefont {Bronson}}, \bibinfo
		{author} {\bibfnamefont {Y.~W.}\ \bibnamefont {Lui}}, \bibinfo {author}
		{\bibfnamefont {C.~M.}\ \bibnamefont {Rozsa}}, \ and\ \bibinfo {author}
		{\bibfnamefont {D.~H.}\ \bibnamefont {Youngblood}},\ }\href {\doibase
		10.1103/physrevlett.45.1670} {\bibfield  {journal} {\bibinfo  {journal}
			{Phys. Rev. Lett.}\ }\textbf {\bibinfo {volume} {45}},\ \bibinfo {pages}
		{1670} (\bibinfo {year} {1980})}\BibitemShut {NoStop}%
	\bibitem [{\citenamefont {YOSHIDA}(2010)}]{Yoshida2010_MPLA25-1783}%
	\BibitemOpen
	\bibfield  {author} {\bibinfo {author} {\bibfnamefont {K.}~\bibnamefont
			{YOSHIDA}},\ }\href {\doibase 10.1142/s0217732310000320} {\bibfield
		{journal} {\bibinfo  {journal} {Mod. Phys. Lett. A}\ }\textbf {\bibinfo
			{volume} {25}},\ \bibinfo {pages} {1783} (\bibinfo {year}
		{2010})}\BibitemShut {NoStop}%
	\bibitem [{\citenamefont {Yoshida}(2010)}]{Yoshida2010_PRC82-034324}%
	\BibitemOpen
	\bibfield  {author} {\bibinfo {author} {\bibfnamefont {K.}~\bibnamefont
			{Yoshida}},\ }\href {\doibase 10.1103/physrevc.82.034324} {\bibfield
		{journal} {\bibinfo  {journal} {Phys. Rev. C}\ }\textbf {\bibinfo {volume}
			{82}},\ \bibinfo {pages} {034324} (\bibinfo {year} {2010})}\BibitemShut
	{NoStop}%
	\bibitem [{\citenamefont {Kvasil}\ \emph {et~al.}(2016)\citenamefont {Kvasil},
		\citenamefont {Nesterenko}, \citenamefont {Repko}, \citenamefont {Kleinig},\
		and\ \citenamefont {Reinhard}}]{Kvasil2016_PRC94-064302}%
	\BibitemOpen
	\bibfield  {author} {\bibinfo {author} {\bibfnamefont {J.}~\bibnamefont
			{Kvasil}}, \bibinfo {author} {\bibfnamefont {V.~O.}\ \bibnamefont
			{Nesterenko}}, \bibinfo {author} {\bibfnamefont {A.}~\bibnamefont {Repko}},
		\bibinfo {author} {\bibfnamefont {W.}~\bibnamefont {Kleinig}}, \ and\
		\bibinfo {author} {\bibfnamefont {P.-G.}\ \bibnamefont {Reinhard}},\ }\href
	{\doibase 10.1103/physrevc.94.064302} {\bibfield  {journal} {\bibinfo
			{journal} {Phys. Rev. C}\ }\textbf {\bibinfo {volume} {94}},\ \bibinfo
		{pages} {064302} (\bibinfo {year} {2016})}\BibitemShut {NoStop}%
	\bibitem [{\citenamefont {Sun}\ and\ \citenamefont
		{Meng}(2022{\natexlab{a}})}]{Sun2022_PRC105-044312}%
	\BibitemOpen
	\bibfield  {author} {\bibinfo {author} {\bibfnamefont {X.}~\bibnamefont
			{Sun}}\ and\ \bibinfo {author} {\bibfnamefont {J.}~\bibnamefont {Meng}},\
	}\href {\doibase 10.1103/physrevc.105.044312} {\bibfield  {journal} {\bibinfo
			{journal} {Phys. Rev. C}\ }\textbf {\bibinfo {volume} {105}},\ \bibinfo
		{pages} {044312} (\bibinfo {year} {2022}{\natexlab{a}})}\BibitemShut
	{NoStop}%
	\bibitem [{\citenamefont {Sun}\ and\ \citenamefont
		{Meng}(2022{\natexlab{b}})}]{Sun2022_PRC106-024334}%
	\BibitemOpen
	\bibfield  {author} {\bibinfo {author} {\bibfnamefont {X.}~\bibnamefont
			{Sun}}\ and\ \bibinfo {author} {\bibfnamefont {J.}~\bibnamefont {Meng}},\
	}\href {\doibase 10.1103/physrevc.106.024334} {\bibfield  {journal} {\bibinfo
			{journal} {Phys. Rev. C}\ }\textbf {\bibinfo {volume} {106}},\ \bibinfo
		{pages} {024334} (\bibinfo {year} {2022}{\natexlab{b}})}\BibitemShut
	{NoStop}%
	\bibitem [{\citenamefont {Washiyama}\ and\ \citenamefont
		{Nakatsukasa}(2017)}]{Washiyama2017_PRC96-041304}%
	\BibitemOpen
	\bibfield  {author} {\bibinfo {author} {\bibfnamefont {K.}~\bibnamefont
			{Washiyama}}\ and\ \bibinfo {author} {\bibfnamefont {T.}~\bibnamefont
			{Nakatsukasa}},\ }\href {\doibase 10.1103/physrevc.96.041304} {\bibfield
		{journal} {\bibinfo  {journal} {Phys Rev C}\ }\textbf {\bibinfo {volume}
			{96}},\ \bibinfo {pages} {041304(R)} (\bibinfo {year} {2017})}\BibitemShut
	{NoStop}%
	\bibitem [{\citenamefont {Shi}\ and\ \citenamefont
		{Stevenson}(2023)}]{Shi2023_CPC47-034105}%
	\BibitemOpen
	\bibfield  {author} {\bibinfo {author} {\bibfnamefont {Y.}~\bibnamefont
			{Shi}}\ and\ \bibinfo {author} {\bibfnamefont {P.~D.}\ \bibnamefont
			{Stevenson}},\ }\href {\doibase 10.1088/1674-1137/acac6b} {\bibfield
		{journal} {\bibinfo  {journal} {Chin. Phys. C}\ }\textbf {\bibinfo {volume}
			{47}},\ \bibinfo {pages} {034105} (\bibinfo {year} {2023})}\BibitemShut
	{NoStop}%
	\bibitem [{\citenamefont {Washiyama}\ \emph {et~al.}(2024)\citenamefont
		{Washiyama}, \citenamefont {Ebata},\ and\ \citenamefont
		{Yoshida}}]{Washiyama2024_PRC109-024317}%
	\BibitemOpen
	\bibfield  {author} {\bibinfo {author} {\bibfnamefont {K.}~\bibnamefont
			{Washiyama}}, \bibinfo {author} {\bibfnamefont {S.}~\bibnamefont {Ebata}}, \
		and\ \bibinfo {author} {\bibfnamefont {K.}~\bibnamefont {Yoshida}},\ }\href
	{\doibase 10.1103/physrevc.109.024317} {\bibfield  {journal} {\bibinfo
			{journal} {Phys. Rev. C}\ }\textbf {\bibinfo {volume} {109}},\ \bibinfo
		{pages} {024317} (\bibinfo {year} {2024})}\BibitemShut {NoStop}%
	\bibitem [{\citenamefont {Nakatsukasa}\ \emph {et~al.}(2007)\citenamefont
		{Nakatsukasa}, \citenamefont {Inakura},\ and\ \citenamefont
		{Yabana}}]{Nakatsukasa2007_PRC76-024318}%
	\BibitemOpen
	\bibfield  {author} {\bibinfo {author} {\bibfnamefont {T.}~\bibnamefont
			{Nakatsukasa}}, \bibinfo {author} {\bibfnamefont {T.}~\bibnamefont
			{Inakura}}, \ and\ \bibinfo {author} {\bibfnamefont {K.}~\bibnamefont
			{Yabana}},\ }\href {http://link.aps.org/doi/10.1103/PhysRevC.76.024318}
	{\bibfield  {journal} {\bibinfo  {journal} {Phys. Rev. C}\ }\textbf {\bibinfo
			{volume} {76}},\ \bibinfo {pages} {024318} (\bibinfo {year}
		{2007})}\BibitemShut {NoStop}%
	\bibitem [{\citenamefont {Avogadro}\ and\ \citenamefont
		{Nakatsukasa}(2011)}]{Avogadro2011_PRC84-014314}%
	\BibitemOpen
	\bibfield  {author} {\bibinfo {author} {\bibfnamefont {P.}~\bibnamefont
			{Avogadro}}\ and\ \bibinfo {author} {\bibfnamefont {T.}~\bibnamefont
			{Nakatsukasa}},\ }\href {\doibase 10.1103/physrevc.84.014314} {\bibfield
		{journal} {\bibinfo  {journal} {Phys. Rev. C}\ }\textbf {\bibinfo {volume}
			{84}},\ \bibinfo {pages} {014314} (\bibinfo {year} {2011})}\BibitemShut
	{NoStop}%
	\bibitem [{\citenamefont {Nik\v{s}i\'c}\ \emph {et~al.}(2008)\citenamefont
		{Nik\v{s}i\'c}, \citenamefont {Vretenar}, \citenamefont {Lalazissis},\ and\
		\citenamefont {Ring}}]{Niksic2008_PRC77-034302}%
	\BibitemOpen
	\bibfield  {author} {\bibinfo {author} {\bibfnamefont {T.}~\bibnamefont
			{Nik\v{s}i\'c}}, \bibinfo {author} {\bibfnamefont {D.}~\bibnamefont
			{Vretenar}}, \bibinfo {author} {\bibfnamefont {G.~A.}\ \bibnamefont
			{Lalazissis}}, \ and\ \bibinfo {author} {\bibfnamefont {P.}~\bibnamefont
			{Ring}},\ }\href {http://link.aps.org/doi/10.1103/PhysRevC.77.034302}
	{\bibfield  {journal} {\bibinfo  {journal} {Phys. Rev. C}\ }\textbf {\bibinfo
			{volume} {77}},\ \bibinfo {pages} {034302} (\bibinfo {year}
		{2008})}\BibitemShut {NoStop}%
	\bibitem [{\citenamefont {Tian}\ \emph
		{et~al.}(2009{\natexlab{a}})\citenamefont {Tian}, \citenamefont {Ma},\ and\
		\citenamefont {Ring}}]{Tian2009_PLB676-44}%
	\BibitemOpen
	\bibfield  {author} {\bibinfo {author} {\bibfnamefont {Y.}~\bibnamefont
			{Tian}}, \bibinfo {author} {\bibfnamefont {Z.~Y.}\ \bibnamefont {Ma}}, \ and\
		\bibinfo {author} {\bibfnamefont {P.}~\bibnamefont {Ring}},\ }\href
	{http://www.sciencedirect.com/science/article/pii/S0370269309004912}
	{\bibfield  {journal} {\bibinfo  {journal} {Phys. Lett. B}\ }\textbf
		{\bibinfo {volume} {676}},\ \bibinfo {pages} {44} (\bibinfo {year}
		{2009}{\natexlab{a}})}\BibitemShut {NoStop}%
	\bibitem [{\citenamefont {Tian}\ \emph
		{et~al.}(2009{\natexlab{b}})\citenamefont {Tian}, \citenamefont {Ma},\ and\
		\citenamefont {Ring}}]{Tian2009_PRC80-024313}%
	\BibitemOpen
	\bibfield  {author} {\bibinfo {author} {\bibfnamefont {Y.}~\bibnamefont
			{Tian}}, \bibinfo {author} {\bibfnamefont {Z.-Y.}\ \bibnamefont {Ma}}, \ and\
		\bibinfo {author} {\bibfnamefont {P.}~\bibnamefont {Ring}},\ }\href
	{http://link.aps.org/doi/10.1103/PhysRevC.80.024313} {\bibfield  {journal}
		{\bibinfo  {journal} {Phys. Rev. C}\ }\textbf {\bibinfo {volume} {80}},\
		\bibinfo {pages} {024313} (\bibinfo {year} {2009}{\natexlab{b}})}\BibitemShut
	{NoStop}%
	\bibitem [{\citenamefont {Nik\v{s}i\'c}\ \emph {et~al.}(2010)\citenamefont
		{Nik\v{s}i\'c}, \citenamefont {Ring}, \citenamefont {Vretenar}, \citenamefont
		{Tian},\ and\ \citenamefont {Ma}}]{Niksic2010_PRC81-054318}%
	\BibitemOpen
	\bibfield  {author} {\bibinfo {author} {\bibfnamefont {T.}~\bibnamefont
			{Nik\v{s}i\'c}}, \bibinfo {author} {\bibfnamefont {P.}~\bibnamefont {Ring}},
		\bibinfo {author} {\bibfnamefont {D.}~\bibnamefont {Vretenar}}, \bibinfo
		{author} {\bibfnamefont {Y.}~\bibnamefont {Tian}}, \ and\ \bibinfo {author}
		{\bibfnamefont {Z.-Y.}\ \bibnamefont {Ma}},\ }\href
	{http://link.aps.org/doi/10.1103/PhysRevC.81.054318} {\bibfield  {journal}
		{\bibinfo  {journal} {Phys. Rev. C}\ }\textbf {\bibinfo {volume} {81}},\
		\bibinfo {pages} {054318} (\bibinfo {year} {2010})}\BibitemShut {NoStop}%
	\bibitem [{\citenamefont {Bjel{\v{c}}i{\'{c}}}\ and\ \citenamefont
		{Nik{\v{s}}i{\'{c}}}(2020)}]{Bjelcic2020_CPC253-107184}%
	\BibitemOpen
	\bibfield  {author} {\bibinfo {author} {\bibfnamefont {A.}~\bibnamefont
			{Bjel{\v{c}}i{\'{c}}}}\ and\ \bibinfo {author} {\bibfnamefont
			{T.}~\bibnamefont {Nik{\v{s}}i{\'{c}}}},\ }\href {\doibase
		10.1016/j.cpc.2020.107184} {\bibfield  {journal} {\bibinfo  {journal}
			{Comput. Phys. Commun.}\ }\textbf {\bibinfo {volume} {253}},\ \bibinfo
		{pages} {107184} (\bibinfo {year} {2020})}\BibitemShut {NoStop}%
	\bibitem [{\citenamefont {Hinohara}\ \emph {et~al.}(2013)\citenamefont
		{Hinohara}, \citenamefont {Kortelainen},\ and\ \citenamefont
		{Nazarewicz}}]{Hinohara2013_PRC87-064309}%
	\BibitemOpen
	\bibfield  {author} {\bibinfo {author} {\bibfnamefont {N.}~\bibnamefont
			{Hinohara}}, \bibinfo {author} {\bibfnamefont {M.}~\bibnamefont
			{Kortelainen}}, \ and\ \bibinfo {author} {\bibfnamefont {W.}~\bibnamefont
			{Nazarewicz}},\ }\href {https://link.aps.org/doi/10.1103/PhysRevC.87.064309}
	{\bibfield  {journal} {\bibinfo  {journal} {Phys. Rev. C}\ }\textbf {\bibinfo
			{volume} {87}},\ \bibinfo {pages} {064309} (\bibinfo {year}
		{2013})}\BibitemShut {NoStop}%
	\bibitem [{\citenamefont {Niksic}\ \emph {et~al.}(2013)\citenamefont {Niksic},
		\citenamefont {Kralj}, \citenamefont {Tutis}, \citenamefont {Vretenar},\ and\
		\citenamefont {Ring}}]{Niksic2013_PRC88-044327}%
	\BibitemOpen
	\bibfield  {author} {\bibinfo {author} {\bibfnamefont {T.}~\bibnamefont
			{Niksic}}, \bibinfo {author} {\bibfnamefont {N.}~\bibnamefont {Kralj}},
		\bibinfo {author} {\bibfnamefont {T.}~\bibnamefont {Tutis}}, \bibinfo
		{author} {\bibfnamefont {D.}~\bibnamefont {Vretenar}}, \ and\ \bibinfo
		{author} {\bibfnamefont {P.}~\bibnamefont {Ring}},\ }\href
	{https://link.aps.org/doi/10.1103/PhysRevC.88.044327} {\bibfield  {journal}
		{\bibinfo  {journal} {Phys. Rev. C}\ }\textbf {\bibinfo {volume} {88}},\
		\bibinfo {pages} {044327} (\bibinfo {year} {2013})}\BibitemShut {NoStop}%
\end{thebibliography}


\end{document}